\documentclass[%
 amssymb, amsmath,%
 aip,cha, table,%
]{revtex4-1}

\usepackage{bm}%
\usepackage{graphicx}
\usepackage{graphics}
\usepackage{amsmath}
\usepackage[table,x11names]{xcolor}
\usepackage{tabularx,colortbl}
\usepackage[colorlinks=true,linkcolor=blue]{hyperref}%
\usepackage{pgfplots}
\pgfplotsset{compat=1.14}
\expandafter\ifx\csname package@font\endcsname\relax\else
 \expandafter\expandafter
 \expandafter\usepackage
 \expandafter\expandafter
 \expandafter{\csname package@font\endcsname}%
\fi
\begin{document} 

\title{Quantifying the Temporal Uncertainties of Nonlinear Turbulence Simulations}

\author{P. Vaezi}
\affiliation{Center for Energy Research, University of California at San Diego, San Diego, California 92093, USA} 
\author{C. Holland}
\affiliation{Center for Energy Research, University of California at San Diego, San Diego, California 92093, USA} 

\begin{abstract}
Nonlinear initial value turbulence simulations often exhibit large temporal variations in their dynamics. Quantifying the temporal uncertainty of turbulence simulation outputs is an important component of validating the simulation results against the experimental measurements, as well as for code-code comparisons. This paper assesses different methods of uncertainty quantification of temporally varying simulated quantities previously used within plasma turbulence community, to evaluate their strengths and potential pitfalls. The use of Autoregressive Moving-Average (ARMA) models for forecasting the uncertainty of turbulence quantities at later simulation times is also studied. These discussions are framed in the practical context of calculating the time-averaging uncertainties of turbulent energy fluxes calculated via gyrokinetic simulations. Particular attention is paid to how standard approaches are challenged as the driving gradient is reduced to the critical value for instability onset.
\end{abstract}

\maketitle


\section{Introduction}

Simulated fluctuation amplitudes and turbulent fluxes often exhibit chaotic behavior, with no obvious regular period or amplitude, and may exhibit large variances and skewnesses. Unlike sampling rates for experimental measurements, where the frequency of measurements combined with sufficient collection windows can often provide enough temporal samples to accurately infer the mean distribution of the quantity, obtaining such long time series in the physics rich simulations is often computationally prohibitive. However, rigorous assessments of simulated quantity statistics, especially their mean value and its uncertainty, is essential for verification and validation (V\&V) studies\cite{holland2016}. Therefore, accurate and computationally feasible estimates of minimum turbulence simulation length are necessary for meaningful V\&V studies. Running simulations long enough to shrink down the uncertainty of the simulation quantity is a favorable endeavor, as it minimizes the temporal uncertainties with respect to other uncertainties, such as uncertainty in fitting profiles to the experimental measurements\cite{chilenski2015}, or input parameter uncertainties into a computational model due to experimental measurements errors\cite{vaezi2018,vaezi2018b}. 

Among temporal uncertainties, determining the variance of the mean distribution for a simulated quantity (in this paper frequently referred as mean variance) is of significant importance. The variance of the mean distribution is needed for determining the temporal uncertainty in V\&V studies and computing the fractional uncertainty of the predicted turbulence quantity\cite{holland2016}. Moreover, advanced reduced models of turbulent plasma transport such as the trapped-gyro-Landau-fluid (TGLF)\cite{staebler2007} model are calibrated to the results of nonlinear simulations. It is therefore essential to ensure uncertainties in the time averaged turbulence levels are small.

Different techniques to measure the fractional uncertainty of nonlinear turbulence simulation have been previously pursued within the magnetic confinement based fusion energy (MFE) research community\cite{mikkelsen2008,holland2016,anderson2017,parker2018}. However, to our knowledge, no rigorous review of measuring and forecasting of temporal fractional uncertainty has been performed within the MFE community. Hence, in this paper we first review the previous methods used within the community to address this issue. Second, we compare the previously used methods of mean variance measurement against the analytic results for model time series to study the pros and cons each method. Third, we examine the convergence of turbulent energy flux means in nonlinear gyrokinetic simulations and forecast temporal fractional uncertainty of turbulence quantities at later simulation times. To carry out this analysis, we use Autoregressive Moving-Average (ARMA) model to forecast the variance of the mean distribution at later simulation times. In practice, we apply the mean variance techniques to gyrokinetic simulation cases in both near and well-above critical temperature gradients.

In this paper, our approach is to discuss practical mean variance measurement techniques without delving too much into their detailed statistics. However, references for more detailed statistical explanations are given for interested readers. In Sec. \ref{sec:meanvaroldtechniques}, we review the previous methods of measuring mean variance, including the integral correlation time method, and sub-interval averaging of correlated measurements. In Sec. \ref{sec:armatest}, we variance results of these previously used techniques against analytic results calculated for ARMA model time series. In Sec. \ref{simulationvar}, we test ARMA model fits to gyrokinetic simulations, to forecast the mean variance at later simulation times and determine the minimum length of simulation required to achieve a desired variance of the mean. In Sec. \ref{summary}, we summarize our study and propose future directions for investigation.

\section{Historical Methods of Measuring Mean Variance of Autocorrelated Measurements} \label{sec:meanvaroldtechniques}

\begin{figure}[ht]
    \centering
    \includegraphics[width=0.37\textwidth]{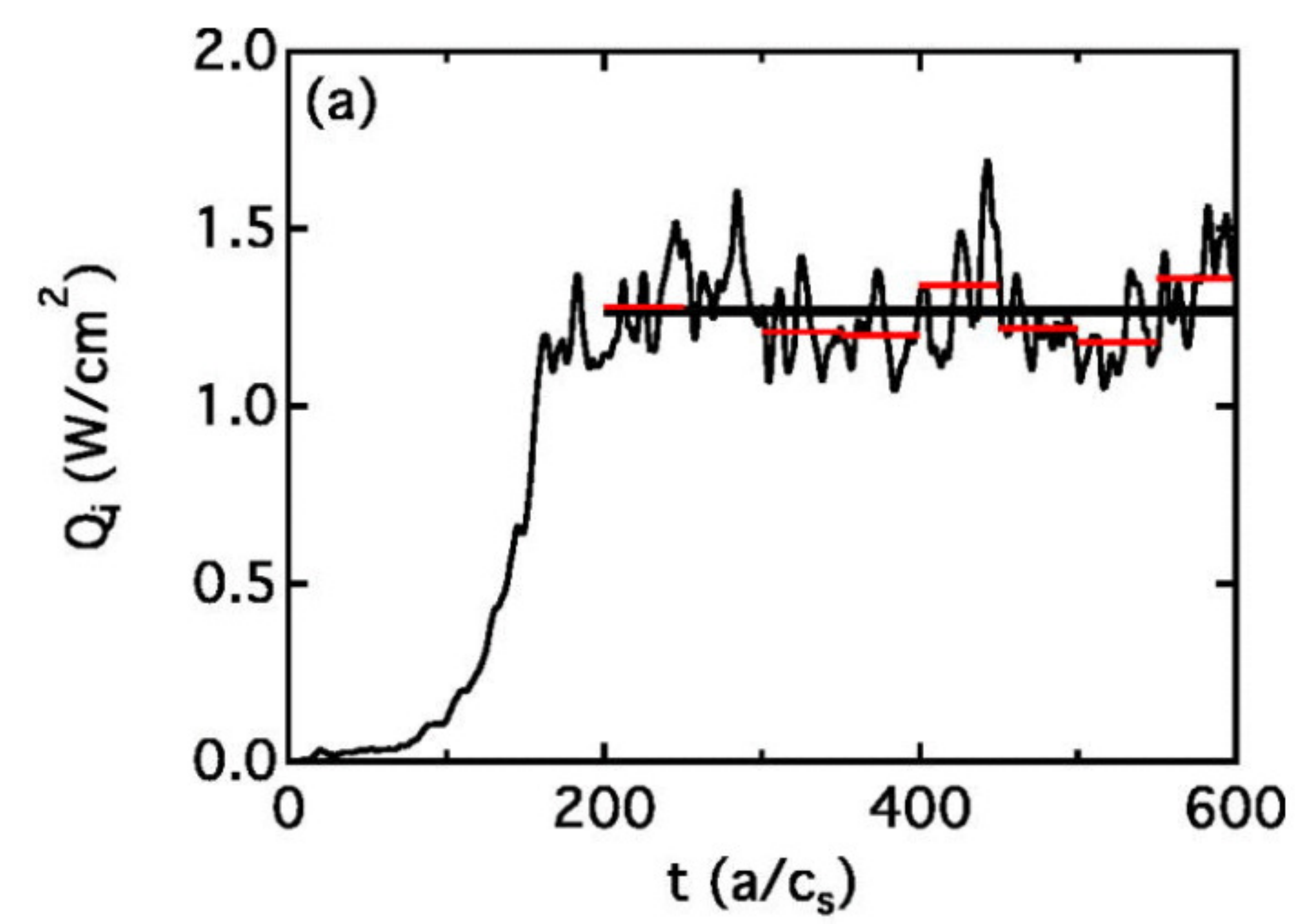}
    \caption{Time trace of a sample ion energy flux from a gyrokinetic turbulence simulation. Reprinted with permission from Holland\cite{holland2016}. Copyright 2016 American Institute of Physics.}
    \label{fig:simtimeseries}
\end{figure}

Calculating the natural variation of turbulence levels within nonlinear initial value simulations is essential to quantifying the temporal uncertainty of turbulence quantities, as shown in Fig. \ref{fig:simtimeseries}. Determining the mean of a turbulence quantity in the saturation phase is straightforward, but determining its aleatory uncertainty and variance is more complicated. In many cases, determining the uncertainty of a mean turbulence quantity in the nonlinear simulations is necessary for verification and validation purposes, as well as calibration of reduced models. In this section, we review some methods of temporal uncertainty estimation used within the MFE turbulence community, and compare these techniques with analytical solutions of the mean distribution variance for model time series.

The simplest possible estimate of the fractional temporal uncertainty of measured quantity is given by
\begin{equation}
\delta_{X} = {\frac{Std[\bar{X}]}{E[\bar{X}]}} =  {\frac{\sqrt{Var[\bar{X}]}}{E[\bar{X}]}},
\end{equation}
where $Std[\bar{X}]$ is the standard deviation of samples $X$, $Var[\bar{X}]$ is the variance of  samples $X$, and $E[\bar{X}]$ is the expected value or the mean of samples $X$.

When the measurement samples are not autocorrelated, the variance of the sample mean of a quantity ${X}$ according to Levy-Lindberg Central Limit Theorem\cite{brown1971} is calculated as
\begin{equation} \label{noncorrvar}
Var[\bar{X}] = {\frac{\sigma_X^2}{n}},
\end{equation}
where $\sigma^2_X = {1/n} \sum_{i=1}^n (X_i - \bar{X})^2$ is the standard deviation of sample, $\bar{X}$ is the mean of samples, and $n$ is the number of repeated measurements. As is well-known, the variance of the mean sampling distribution shrinks down with the availability of more samples. However, if the measurements are correlated, due to for example a high  sampling rate relative to physical timescales, the calculation of the variance of $\bar{X}$ becomes more complicated\cite{zhang2006}. For measurements with finite autocorrelation, Andrews\cite{andrews1991} suggests estimating the long-run variance using a kernel estimator,
\begin{equation}\label{corrvar}
\hat{\sigma}_X^2 = \sum_{|j| \leq q} \omega_q(j) \rho_{X}(j);
\end{equation}
\begin{equation} \rho_X(j) = {\frac{1}{n}} \sum_{i=1}^{n - |j|} \left(X_i - \bar{X}\right) \left(X_{i+|j|} - \bar{X} \right).
\end{equation}
In the above equations, $\hat{\sigma}_X^2$ is the variance of time series accounting for autocorrelated lags, $q$ is a specified cut-off lag, and $\omega_q(j)$ is a symmetric, bounded, and integrable kernel estimator. The kernel estimator mathematically is a symmetric filtering function to filter out the noise of larger lags arising from finite length of sampled time series. Possible choices for the kernel estimator include the Barlett kernel, truncated kernel, and Hanning kernel\cite{anderson2011}. If measurements become uncorrelated, the covariance terms become zero and $\rho_X(j)$ reduces to the standard deviation of the samples, recovering Eqn. \ref{noncorrvar}.

To approximate the mean variance of autocorrelated processes, different heuristic approaches have been used within the MFE community. Two examples considered here are
\begin{enumerate}
\item Using an approximate integral correlation time in the calculation of the variance.
\item Splitting the signal into sub-windows of sufficient length that the sample sub-window means are no longer correlated, and calculating the variance of uncorrelated sub-window means will become an approximation of the mean variance of original time series.
\end{enumerate}

In the following subsections, we describe these two different approaches, and in Sec. \ref{sec:armatest} we compare the variance calculation of these approaches against analytical values of the mean variance for model time series. We note that the methods described and pursued in this manuscript assumes that the time series is stationary and ergodic, therefore can only be applied to the saturated phase of simulations. Possible approaches for future studies which relax this restriction are discussed in Sec. \ref{summary}.

\subsection{Integral Correlation Time}

In this approach, the variance of $\bar{X}$ is originated from Eqn. \ref{corrvar} by assuming each time step is a measurement sample, and the correction coefficient due to finite autocorrelation of the measurements is approximated as
\begin{equation} \label{ictmethod}
Var[\bar{X}] = {\tau_{int}} {\frac{\sigma_X^2}{n}}.
\end{equation}

Essentially, the effective number of samples is reduced by the integral correlation time $\tau_{int}$. The problem is now reduced to estimating the integral correlation time $\tau_{int}$ over all the time lags. However, due to finite sample size of our measurement, larger lag may pose error on the estimate of $\tau_{int}$. A practical estimate of $\tau_{int}$ is proposed by Nevins\cite{nevins2004} is
\begin{equation}
\tau_{int} = \int_{-\tau_{lag}}^{\tau_{lag}} d\tau \bar{C}_{X}(\tau),
\end{equation}
where the $\tau_{lag} \approx \sqrt{\tau_c T}$, $\tau_c$ is the width of the lag region in which autocorrelation is above standard error
\begin{equation} \label{standarderror}
S.E.(\tau) = \sqrt{{\frac{1}{N}}\left(1 + 2 \sum_{i=1}^{\tau} \rho_i^2 \right)},
\end{equation}
and $\bar{C}_{X}(\tau)$ is the auto-variance of time lag $\tau$ passed into a Hanning kernel which can be obtained as,
\begin{equation}
\bar{C}_{X}(\tau) = \mathbb{H}(\tau / \tau_{lag}) C_{X}(\tau),
\end{equation}
in which $\mathbb{H}$ is the Hanning function, and $C_{X}(\tau)$ is the standard estimate of the auto-variance,
\begin{equation}
C_{X}(\tau) = {\frac{1}{T}} \int_{0}^{T - \tau} dt X(t) X(t+\tau).
\end{equation}

Hence, in practice one can calculate auto-variance of different time lags, pass the auto-variance function through a symmetric kernel estimator (here a Hanning function is used), and integrate the weighted auto-variance function over significant time lags. If there is no correlated lag, the integral correlation time reduces to one, retaining the uncorrelated variance of samples.

\begin{figure}[ht]
    \centering
    \includegraphics[trim={0 0.5cm 0 0},clip,width=0.4\textwidth]{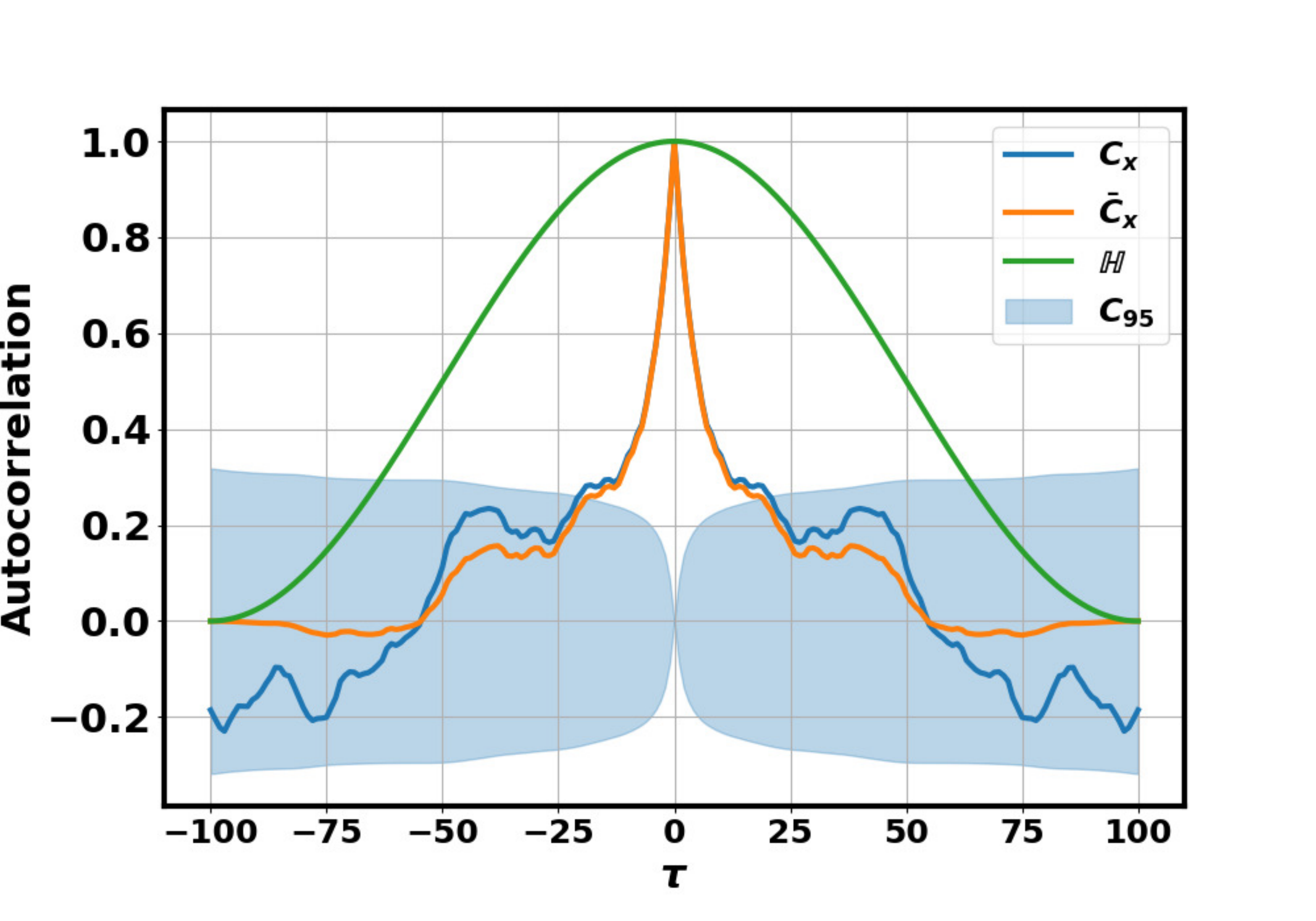}
    \caption{Autocorrelation function shown with the shaded \%95 confidence intervals ($\pm 1.96\sqrt{S.E.}$), as well as fitted Hanning kernel estimator for an arbitrary time series of 500 samples. {The analysis shown here is taken from time series of Fig. }\ref{fig:subintervalts}.}
    \label{fig:intcorrtime}
\end{figure}

To illustrate integral correlation time variance measurement technique, in Fig. \ref{fig:intcorrtime} we have shown the autocorrelation function of a sample time series with the large-lag standard error. A Hanning kernel filter is passed to remove the noise arising from large-lag terms. The filtered auto-variance is therefore weighted to compute the integral correlation time. A result, the variance of sample can be adjusted to account for the autocorrelated measurements. As might be expected, this mean variance estimatation technique is highly sensitive to the width of lag region, and the choice of kernel estimator.

\subsection{Sub-interval Averaging of Correlated Measurements}

In the second approach\cite{mikkelsen2008b}, the measurement samples are taken as the means of $N$ non-overlapping sub-interval windows of the original time series of length $T$. These subwindow means are denoted with $Y_i$ here and can be obtained as
$$
Y_1 = {X_1 + \dots + \frac{X_{T}}{T}} $$
$$\dots$$
\begin{eqnarray}Y_{N} = {X_{(N-1)T + 1} + \dots + \frac{X_{NT}}{T}}.
\end{eqnarray} 
Here, by selecting large enough sub-interval width $T$, the lagged autocorrelation of sub-window means become statistically insignificant. We can therefore use the subwindow means to estimate the variance of the mean for the total signal via Eqn. \ref{noncorrvar}, which in this case reads as
\begin{equation} \label{subwindowmethod}
Var[\bar{X}] \sim Var[\bar{Y}] = {\frac{\sigma_{{Y}}^2}{N}},
\end{equation}
where $\sigma_Y^2$ is the variance of sub-interval means.

The key question for this approach is how to obtain the minimum acceptable value of $T$. To obtain the minimum sub-interval width for uncorrelated sub-windows, we can test the statistical significance of lagged correlations for sub-window means. Typically, the sub-interval width should be larger than turbulence timescale. For a stochastic process where the autocorrelation exponentially drops, a quick comparison of the first time lag autocorrelation of sub-window means, $\rho_Y(1)$, against their standard error\cite{parzen1963} is sufficient to determine if the width of sub-window averaging is large enough or not. Autocorrelation of the first moving-average lag below standard error can be approximated as uncorrelated moving-averages.

\begin{figure}[ht]
    \centering
    \includegraphics[trim={0 0.5cm 0 0},clip,width=0.4\textwidth]{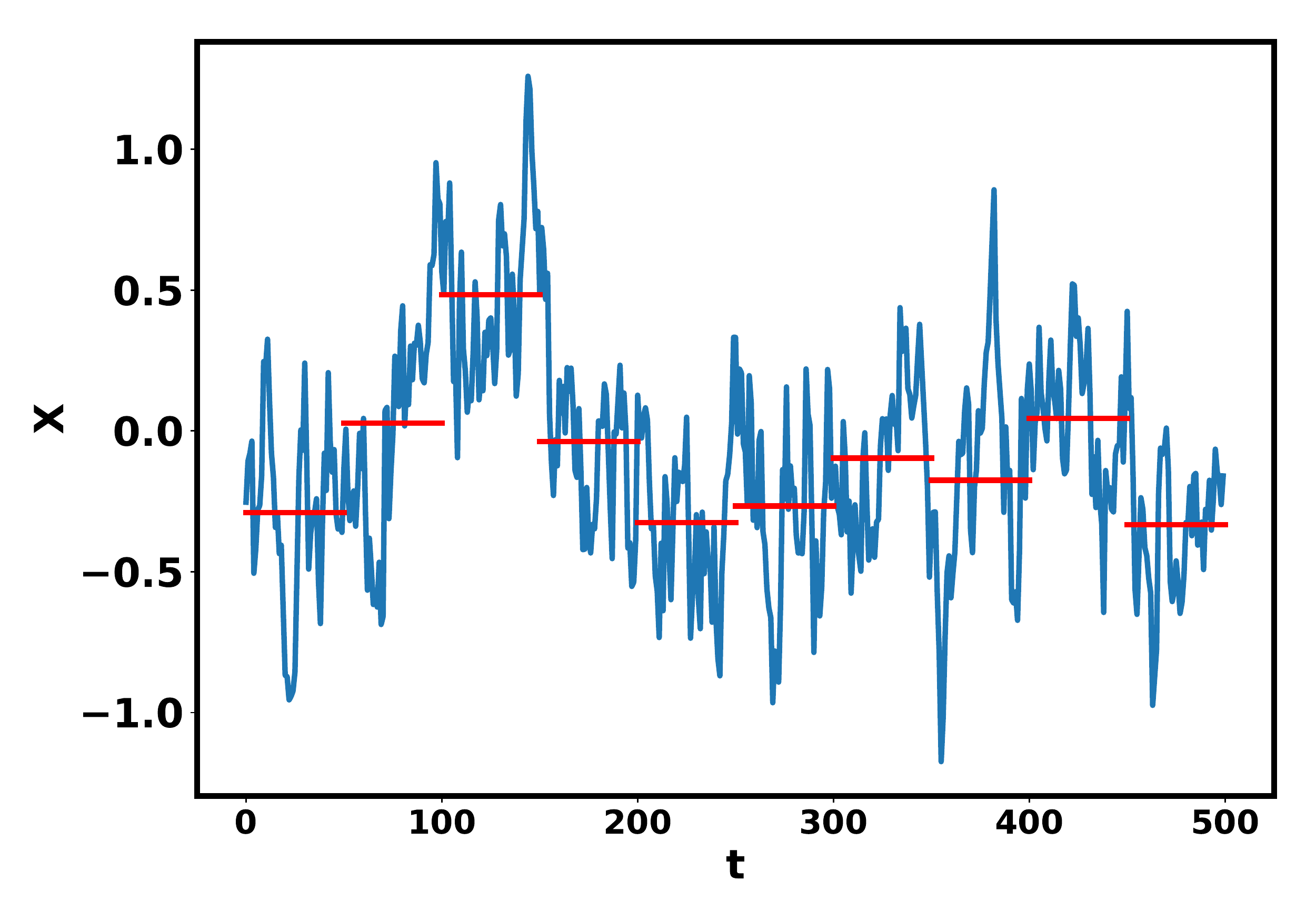}
    \caption{A sample time series with 10 sub-interval means, denoted by the horizontal lines.}
    \label{fig:subintervalts}
\end{figure}

For better illustration of this mean variance estimation technique, in Fig. \ref{fig:subintervalts} we have shown a sample time series with the ten sub-interval means. If the sub-window means have statistically insignificant correlation, we can easily calculate the uncorrelated variance of sub-window means as the mean variance of the time series.

\section{An Analytical Mean Variance Study} \label{sec:armatest}

In this Section, we will examine application of Auto-Regressive Moving Average\cite{choi2012arma} (ARMA) models to analytical time series. ARMA models provide a description of stationary and ergodic processes in terms of two set of polynomials, one for the autoregression and the second for the moving average, and are commonly used in a variety of research communities. The general format of ARMA$(p,q)$ models with $p$ autoregressive terms and $q$ moving-average terms is formulated as
\begin{equation}
X_t = \phi_1 X_{t-1} + \dots + \phi_p X_{t-p} + \varepsilon_t + \theta_1 \varepsilon_{t-1} + \dots + \theta_q \varepsilon_{t-q},
\end{equation}
where $\phi_i$ and $\theta_i$ are parameters of the model, and $\varepsilon_t, \varepsilon_{t-1}, ...$ are white noises with $\mathcal{N}(0, \sigma_\varepsilon^2)$.

Here, we will look at an example of using an ARMA model with one auto-regressive term and zero moving-averages to generate sample time series, and compare the efficacy of the methods introduced in Sec. \ref{sec:meanvaroldtechniques} against the analytical derivation of mean variance.

For the ARMA$(1,0)$ model $X_t = \phi_1 X_{t-1} + \varepsilon_t$, the variance of the sample mean analytically asymptotes to\cite{crack2004}
\begin{equation}
Var[\bar{X}] \sim {\frac{\sigma_X^2 (1 + \phi_1)}{n (1 - \phi_1)}},
\end{equation}
and autocorrelation function of ARMA$(1,0)$ is given by\cite{thompson2010comparison}
\begin{equation}
\rho_X(0) = 1; \rho_X(i) = \left(\phi_1\right)^i.
\end{equation}

\begin{figure*}[!htb]
  \centering
  \begin{minipage}[b]{0.325\textwidth}
    \includegraphics[trim={0cm 0 0 0},clip, width=1.\textwidth]{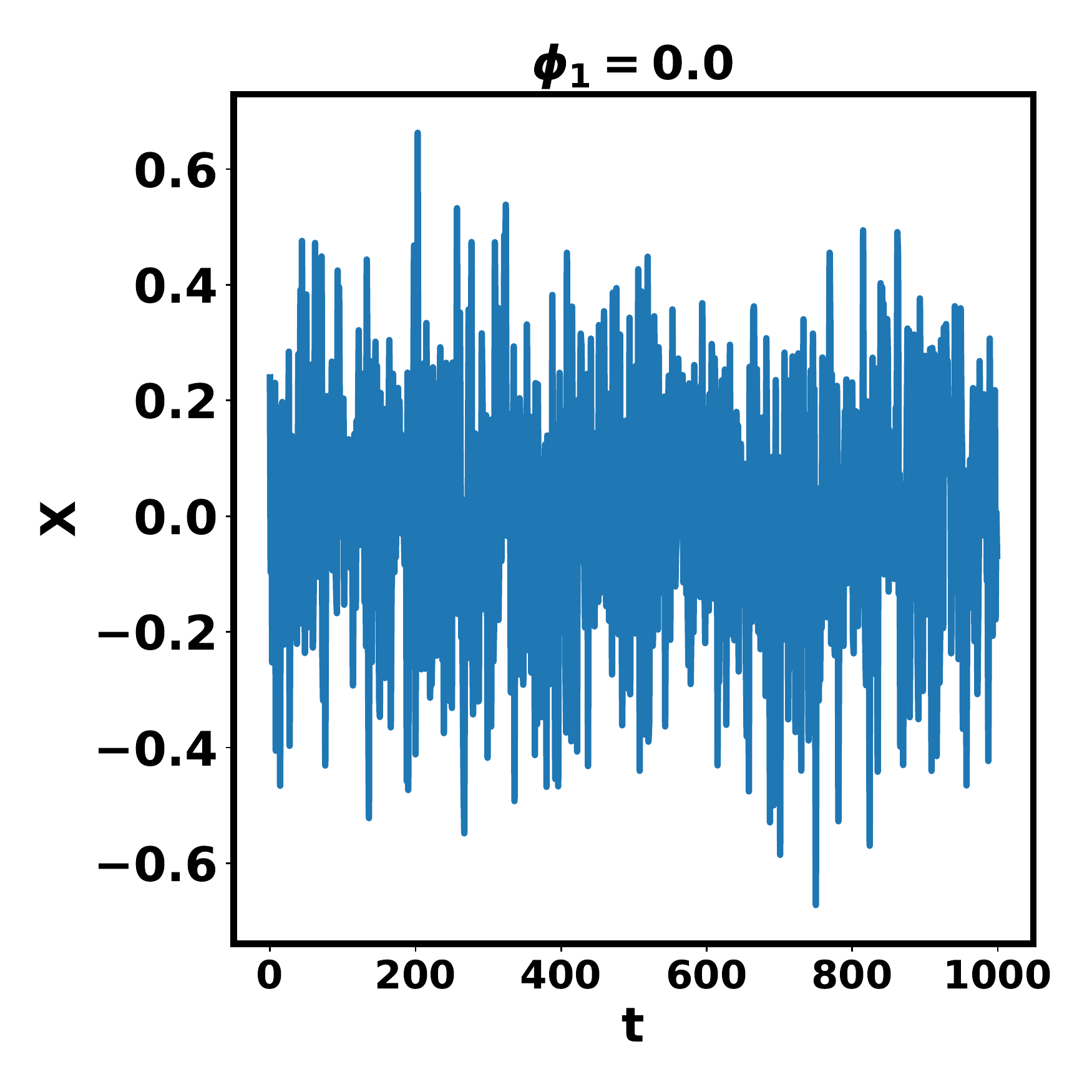}
  \end{minipage}
  \begin{minipage}[b]{0.325\textwidth}
    \includegraphics[trim={0cm 0 0 0},clip, width=1.\textwidth]{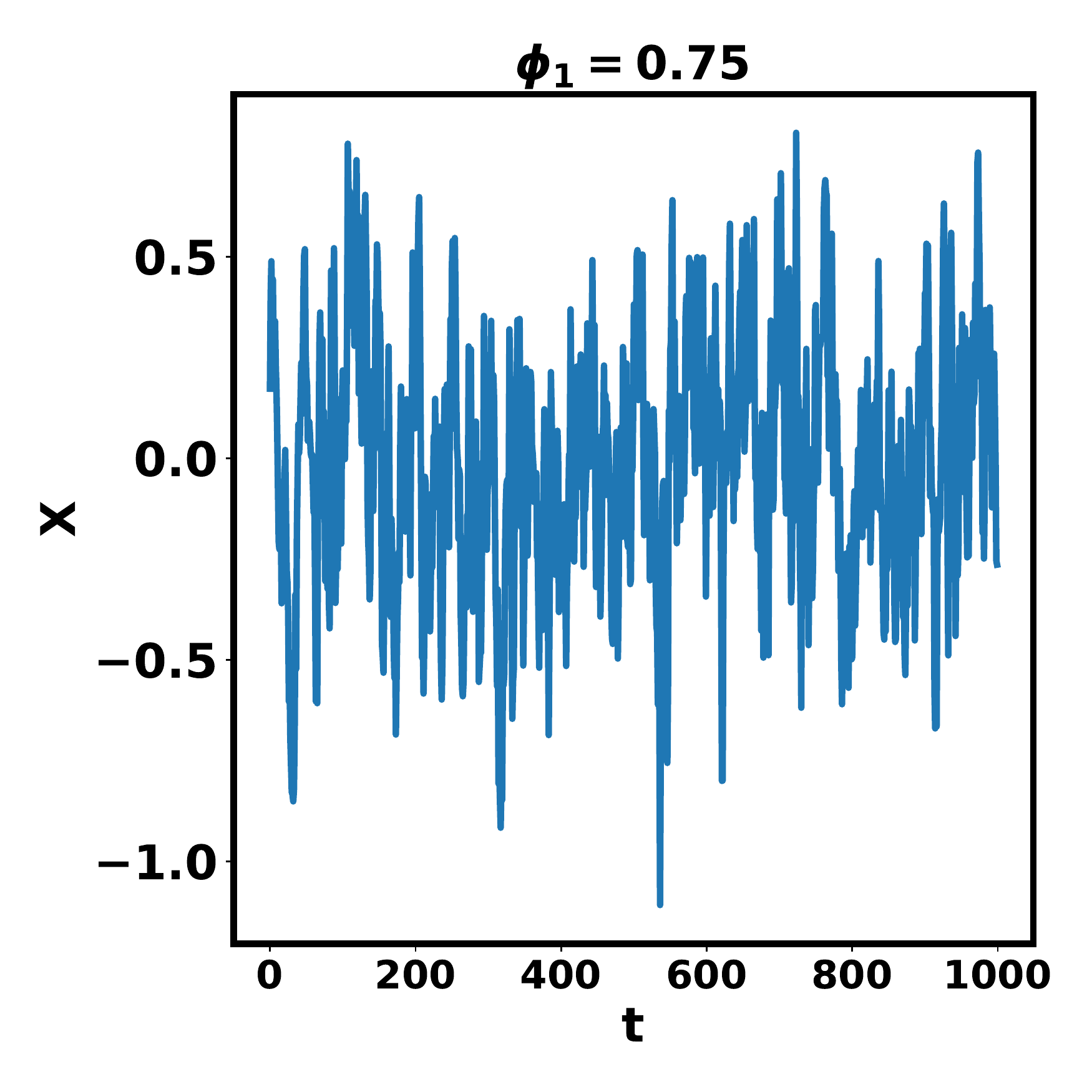}
  \end{minipage}
  \begin{minipage}[b]{0.325\textwidth}
    \includegraphics[trim={0cm 0 0 0},clip, width=1.\textwidth]{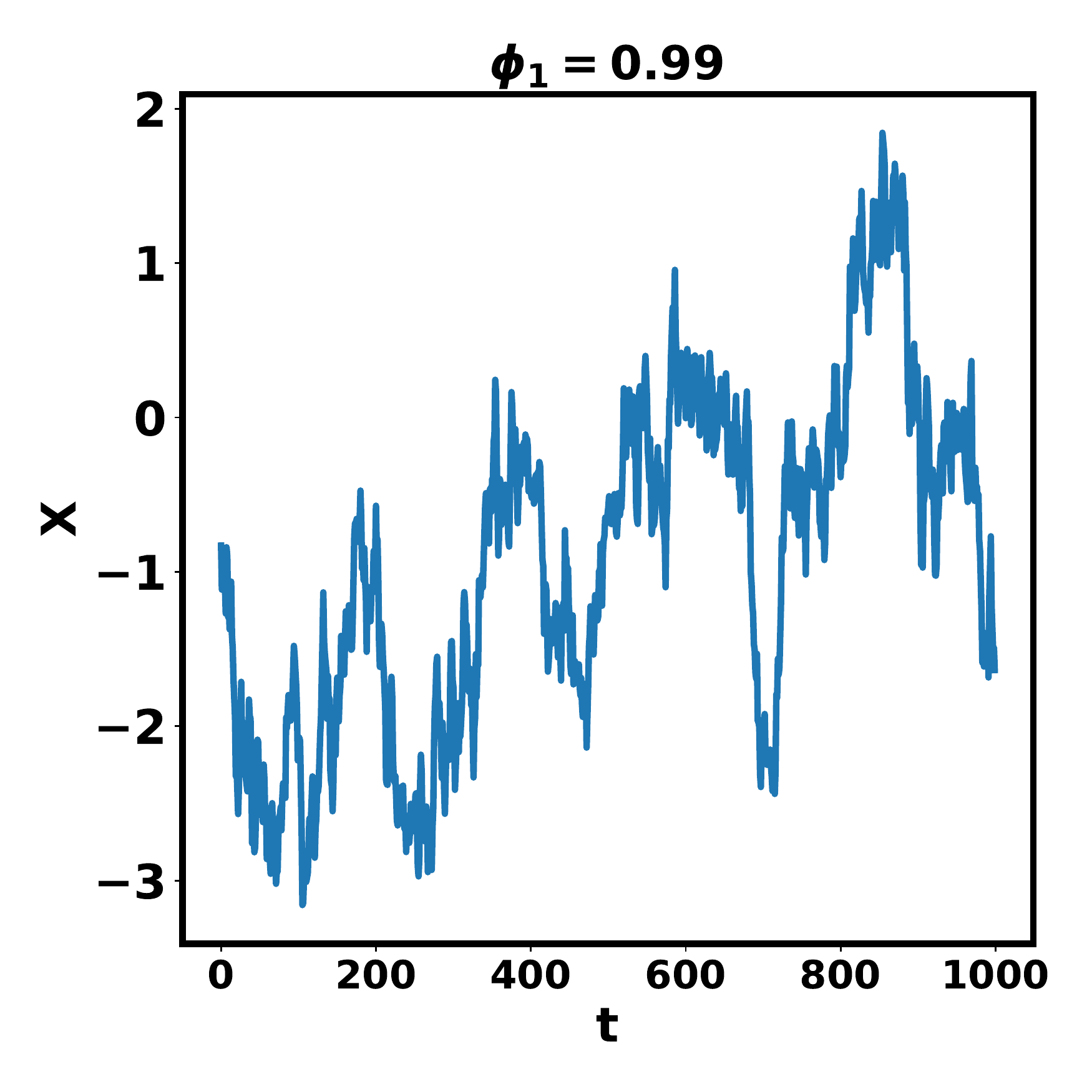}
  \end{minipage}
  \caption{Randomly generated ARMA$(1,0)$ process with 1000 samples for different values of $\phi_1 = \{0, 0.75, 0.99\}$.}
  \label{fig:armats}
\end{figure*}

\begin{figure*}[!htb]
  \centering
  \begin{minipage}[b]{0.325\textwidth}
    \includegraphics[trim={0cm 0 0 0},clip, width=1.\textwidth]{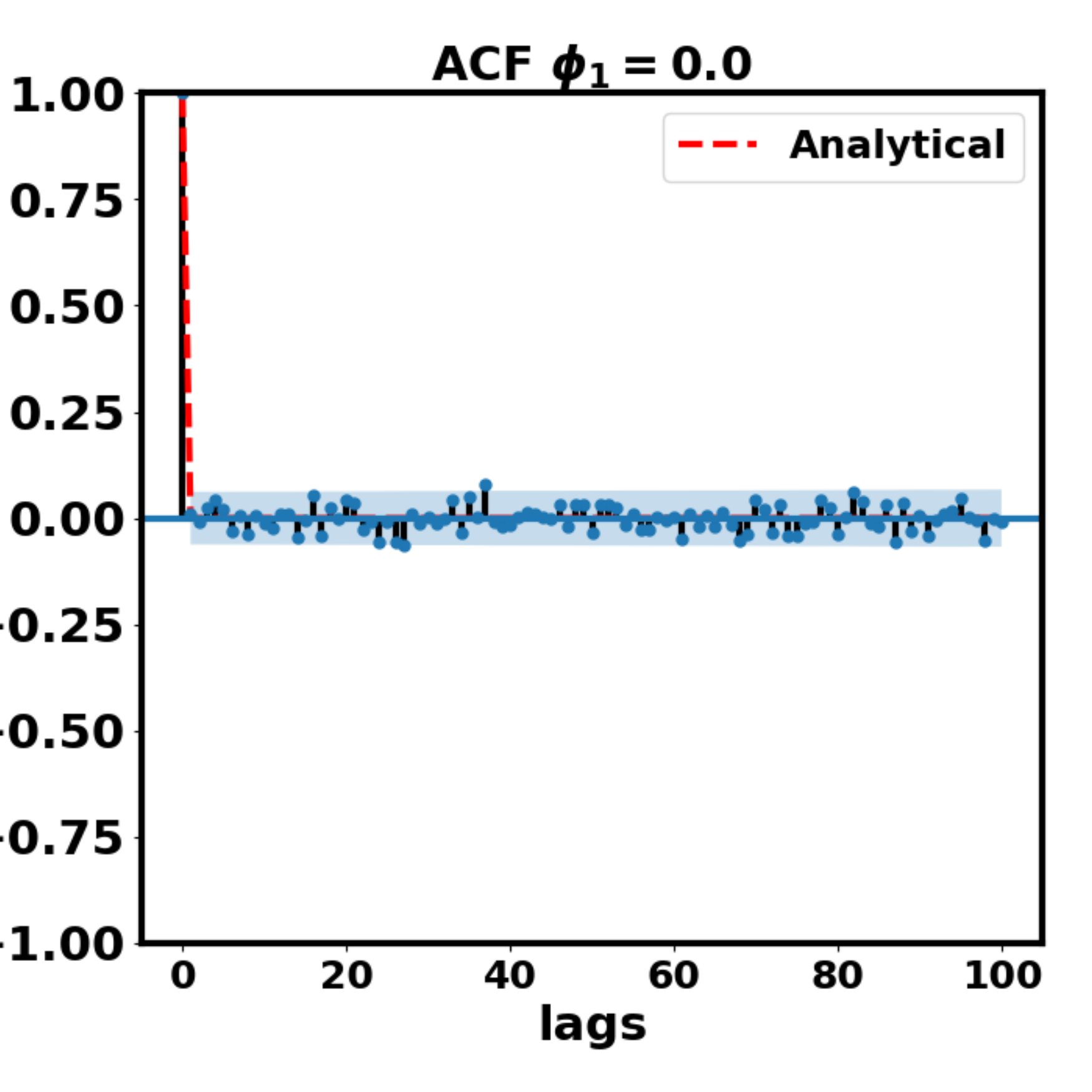}
  \end{minipage}
  \begin{minipage}[b]{0.325\textwidth}
    \includegraphics[trim={0cm 0 0 0},clip, width=1.\textwidth]{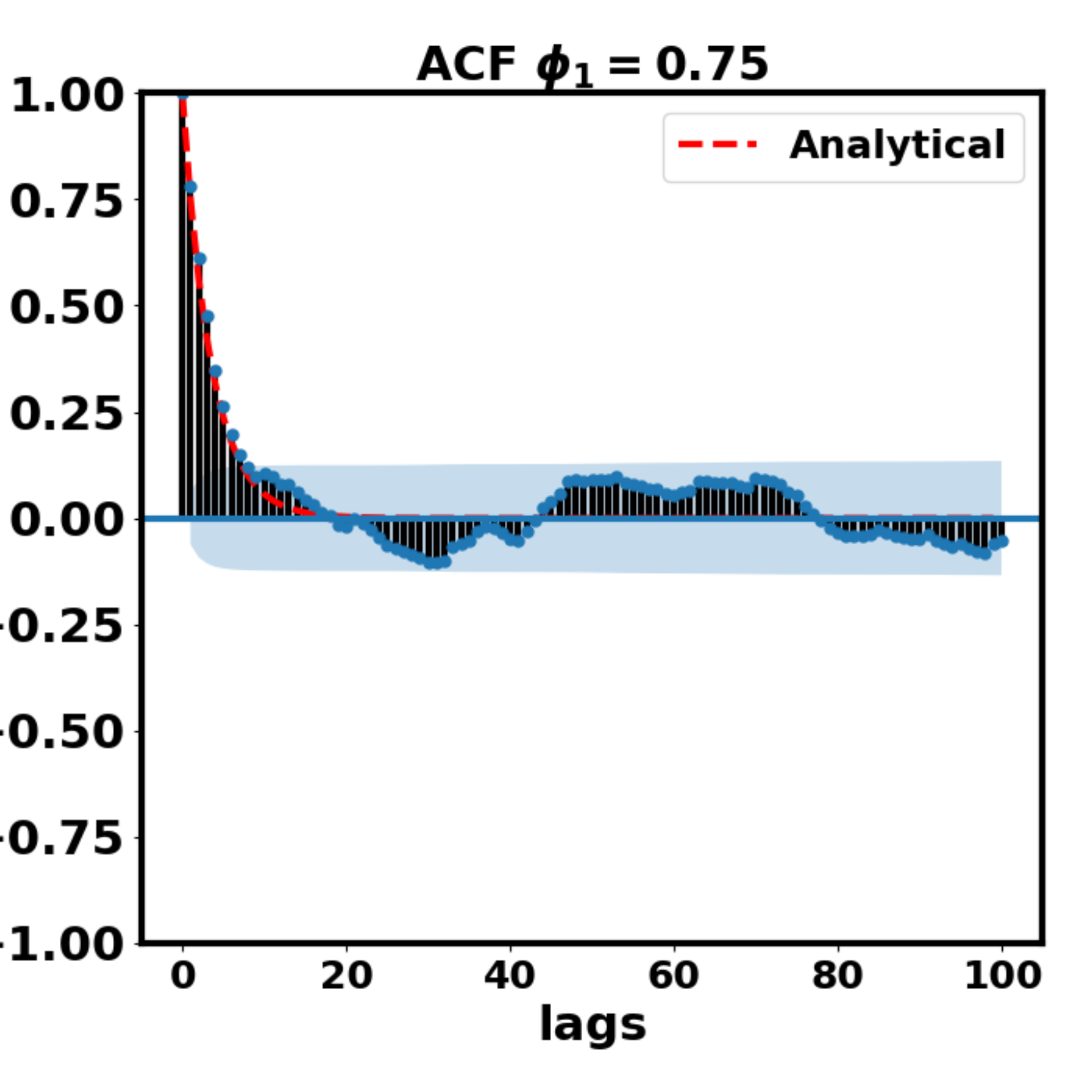}
  \end{minipage}
  \begin{minipage}[b]{0.325\textwidth}
    \includegraphics[trim={0cm 0 0 0},clip, width=1.\textwidth]{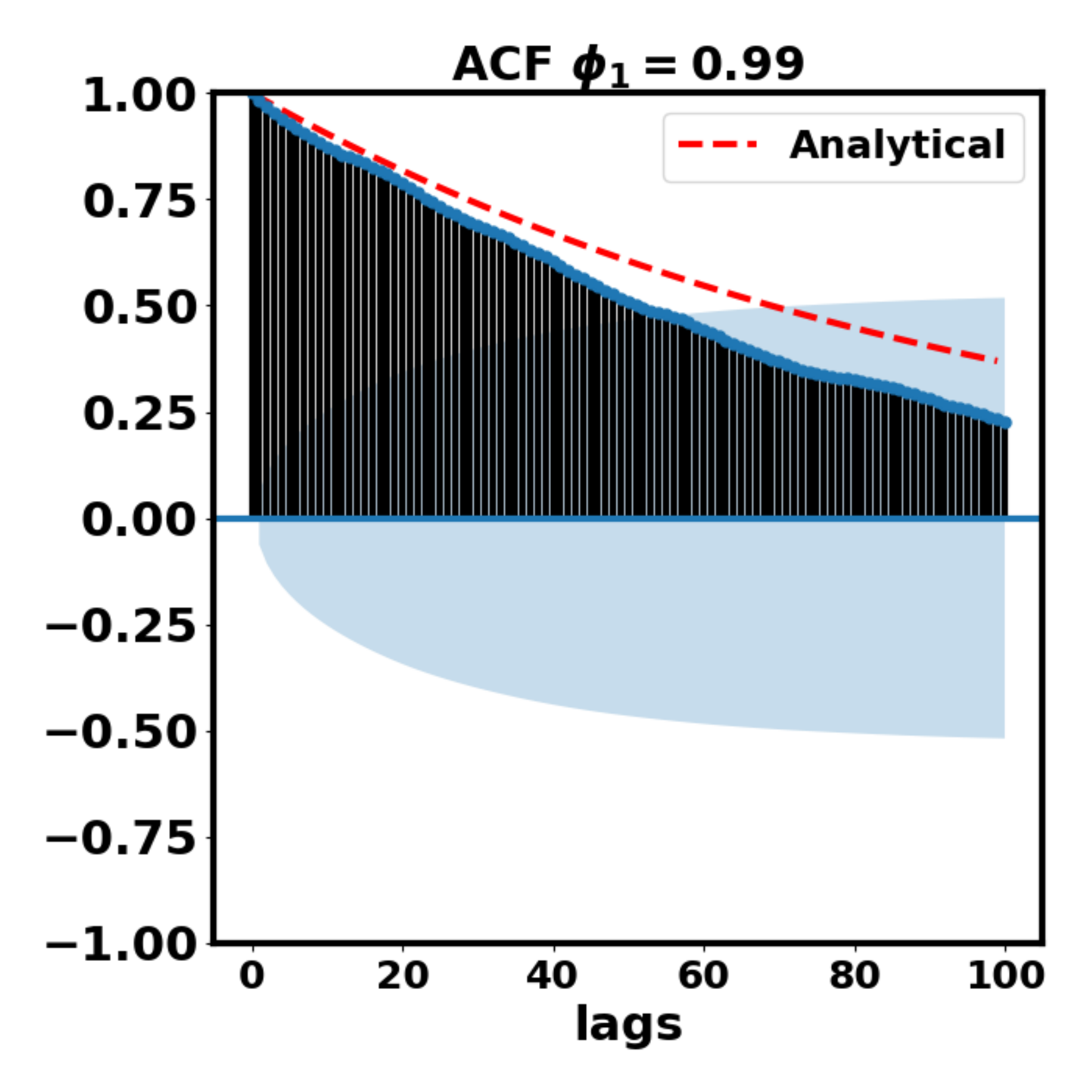}
  \end{minipage}
  \caption{Autocorrelation function of ARMA$(1,0)$ process with 1000 samples for different values of $\phi_1 = \{0, 0.75, 0.99\}$ compared with analytical autocorrelation function of ARMA$(1,0)$ model. The shaded band shows standard error bounds.}
  \label{fig:armaacf}
\end{figure*}

We can observe when $\phi_1$ is finite, the autocorrelation of lags are finite and decreasing with the lag number, and the mean variance is no longer equal to uncorrelated case of Eqn. \ref{noncorrvar}. We now consider three time series generated using ARMA$(1,0)$ model to study the effects of correlation persistence in finite sample size time series samples. In Fig. \ref{fig:armats}, we show a randomly generated finite sample size ($n=1000$ and $\sigma_\varepsilon = 0.2$) ARMA$(1,0)$ process with $\phi_1=\{0,0.75,0.99\}$. In Fig. \ref{fig:armaacf} we plot the time series sample autocorrelation function against the analytical value of autocorrelation for an ARMA$(1,0)$ process. We observe for smaller $\phi_1$, with less significant lags, 1000 samples used are sufficient to converge to analytical autocorrelation function. On the other hand, we observe for larger $\phi_1$ values the autocorrelation persistence is significantly larger, and our finite number of samples are not enough to accurately approximate the analytical autocorrelation function. In such large persistence processes, due to cumulative summation of larger correlation lags, the standard error (Eqn. \ref{standarderror}) is much larger, resulting in much larger confidence interval bounds. We see therefore that larger autocorrelation effects and finite sample number can have drastic influence on the calculation of the mean variance.

To understand which method can better approximate the mean variance of the signal, we have compared mean variance of introduced methods in the previous Section (Eqns. \ref{noncorrvar}, \ref{ictmethod}, and \ref{subwindowmethod}) against the analytical value of mean variance for ARMA$(1,0)$, and results are shown in Table \ref{tbl:compmethods}. We can gain some essential insights from Table \ref{tbl:compmethods} as how these different methods work in different correlation persistence settings.

\begin{figure*}[!htb]
  \centering
  \begin{minipage}[b]{0.325\textwidth}
    \includegraphics[trim={0cm 0 0 0},clip, width=1.\textwidth]{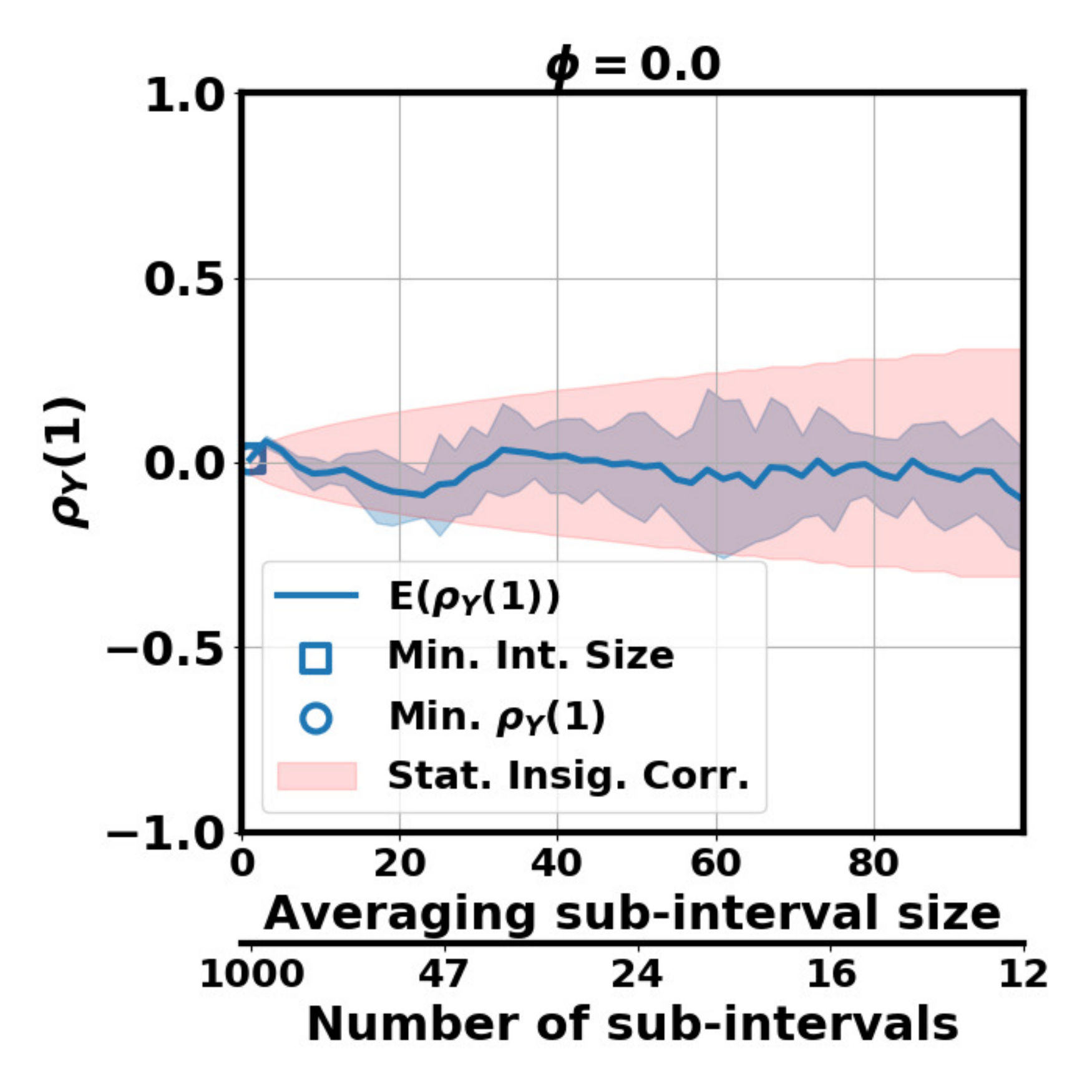}
  \end{minipage}
  \begin{minipage}[b]{0.325\textwidth}
    \includegraphics[trim={0cm 0 0 0},clip, width=1.\textwidth]{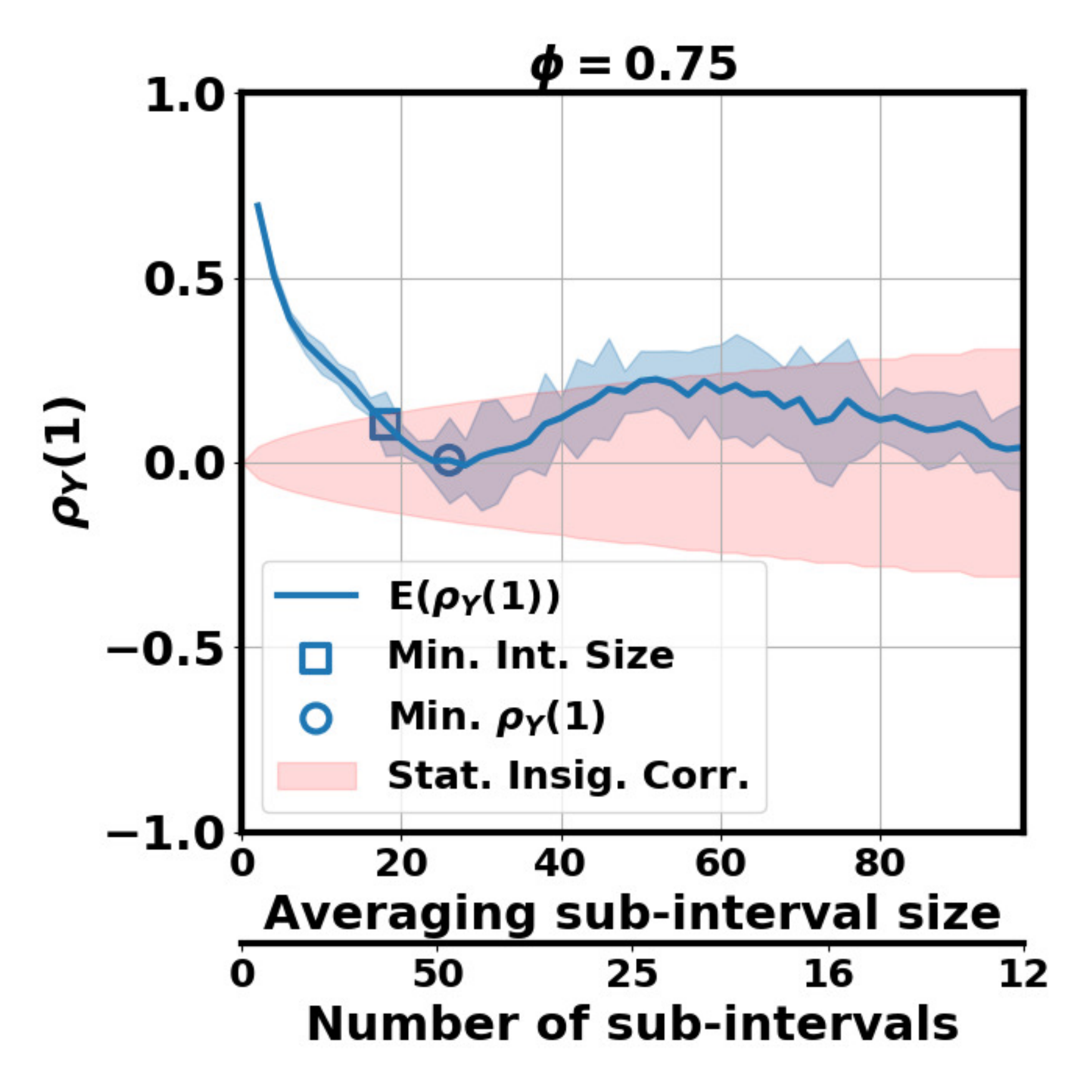}
  \end{minipage}
  \begin{minipage}[b]{0.325\textwidth}
    \includegraphics[trim={0cm 0 0 0},clip, width=1.\textwidth]{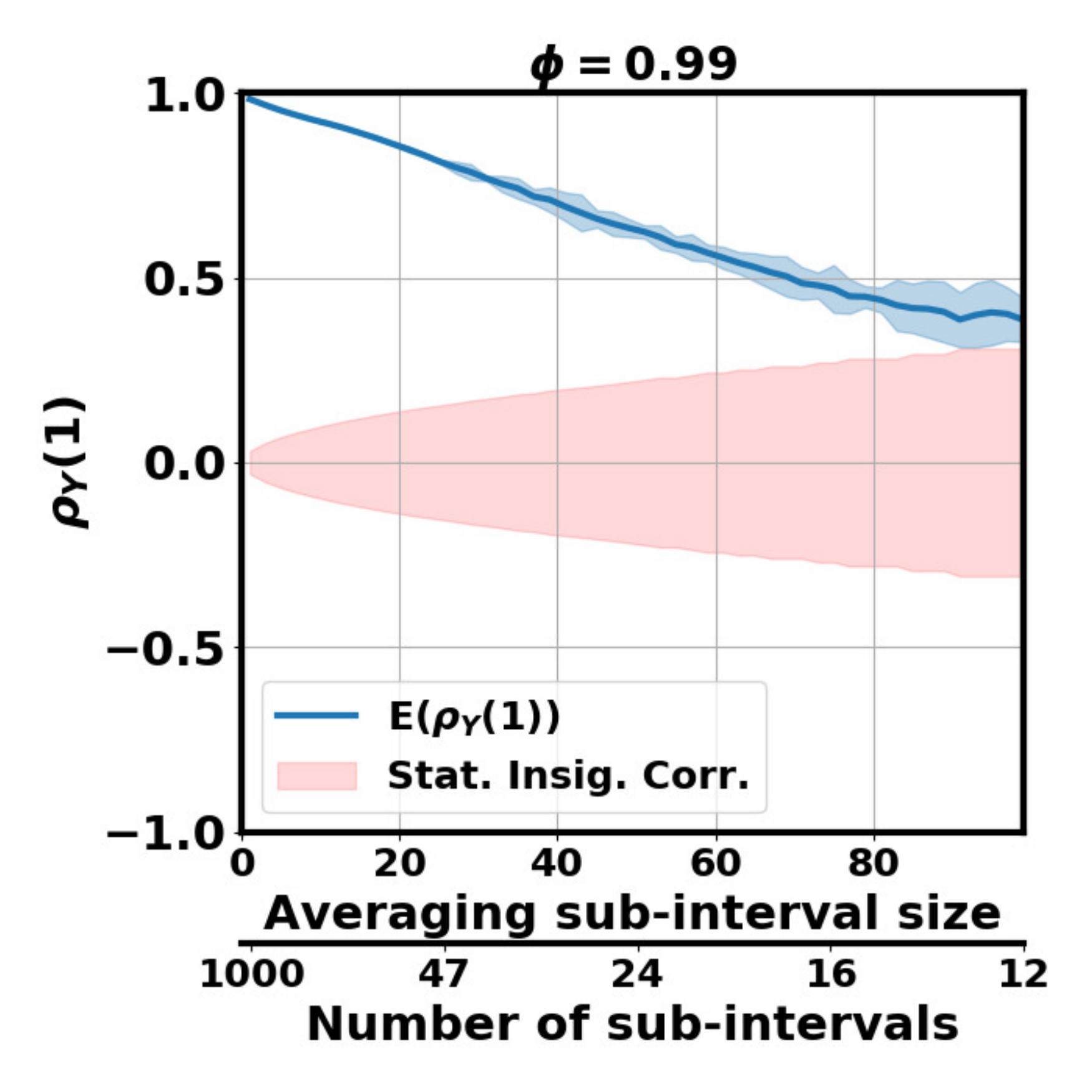}
  \end{minipage}
  \caption{Autocorrelation of first sub-window average lag, $\rho_Y(1)$, of ARMA$(1,0)$ process as a function of sub-window size with 1000 samples for different values of $\phi_1 = \{0, 0.75, 0.99\}$. The zero-symmetric shaded band shows correlation statistical significance rejection area.}
  \label{fig:armarho1}
\end{figure*}

Before we proceed with the comparison, we should note that for the sub-interval average method, we need to determine the size of sub-window. Statistically, due to finite sample size of time series, a correlation of lag below standard error statistically can not reject the uncorrelated sub-intervals hypothesis. In Fig. \ref{fig:armarho1}, we have shown the autocorrelation of sub-interval averaged first lags $\rho_Y(1)$ as function of sub-interval size. The zero symmetric shaded area show the statistically insignificant region where the autocorrelation of first lag is below standard error. Moreover, the line and its shaded area show the expected value and its deviation due to independent $\rho_Y(1)$ calculation of sub-interval shifts.

\begin{figure*}[!htb]
  \centering
  \begin{minipage}[b]{0.325\textwidth}
    \includegraphics[trim={0cm 0 0 0},clip, width=1.\textwidth]{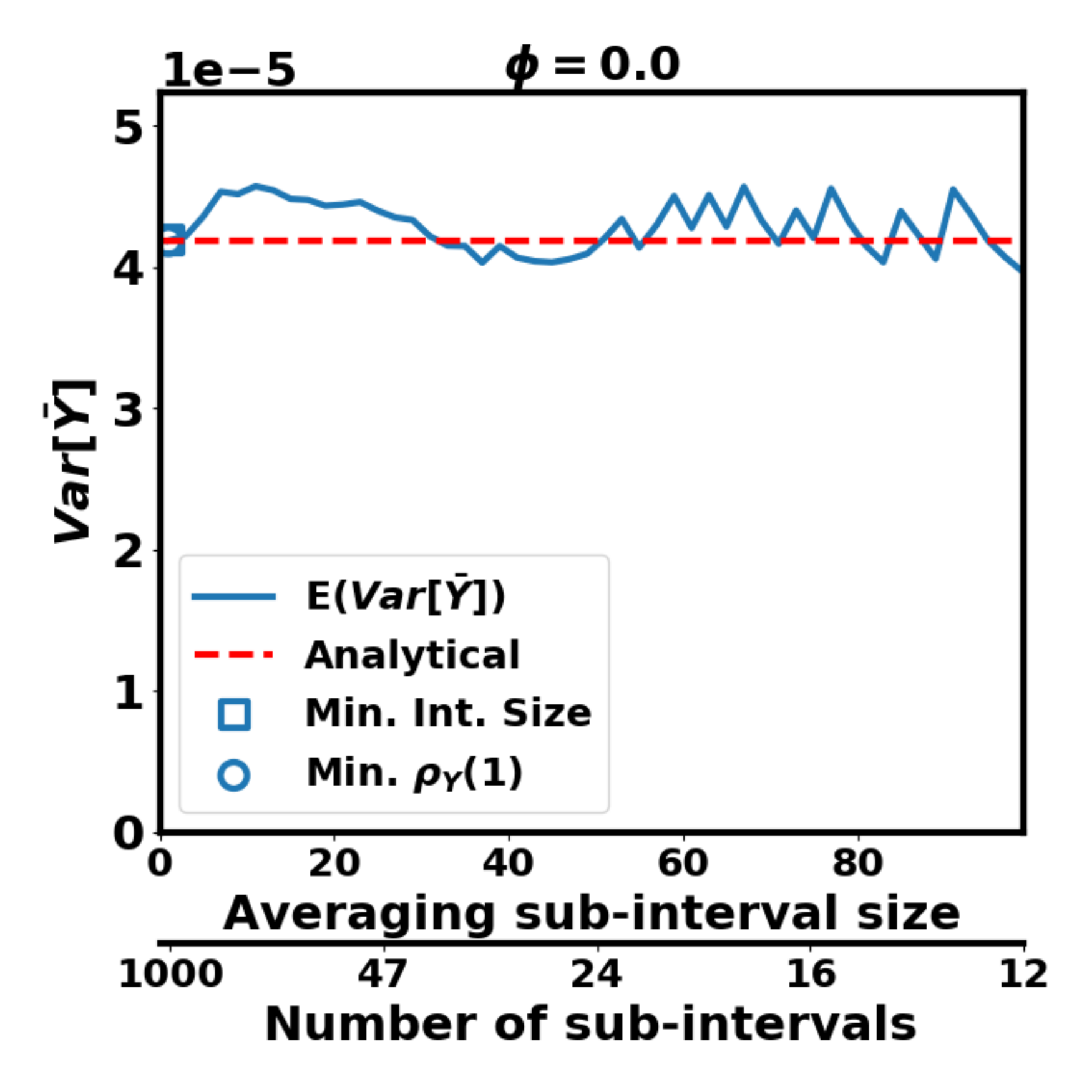}
  \end{minipage}
  \begin{minipage}[b]{0.325\textwidth}
    \includegraphics[trim={0cm 0 0 0},clip, width=1.\textwidth]{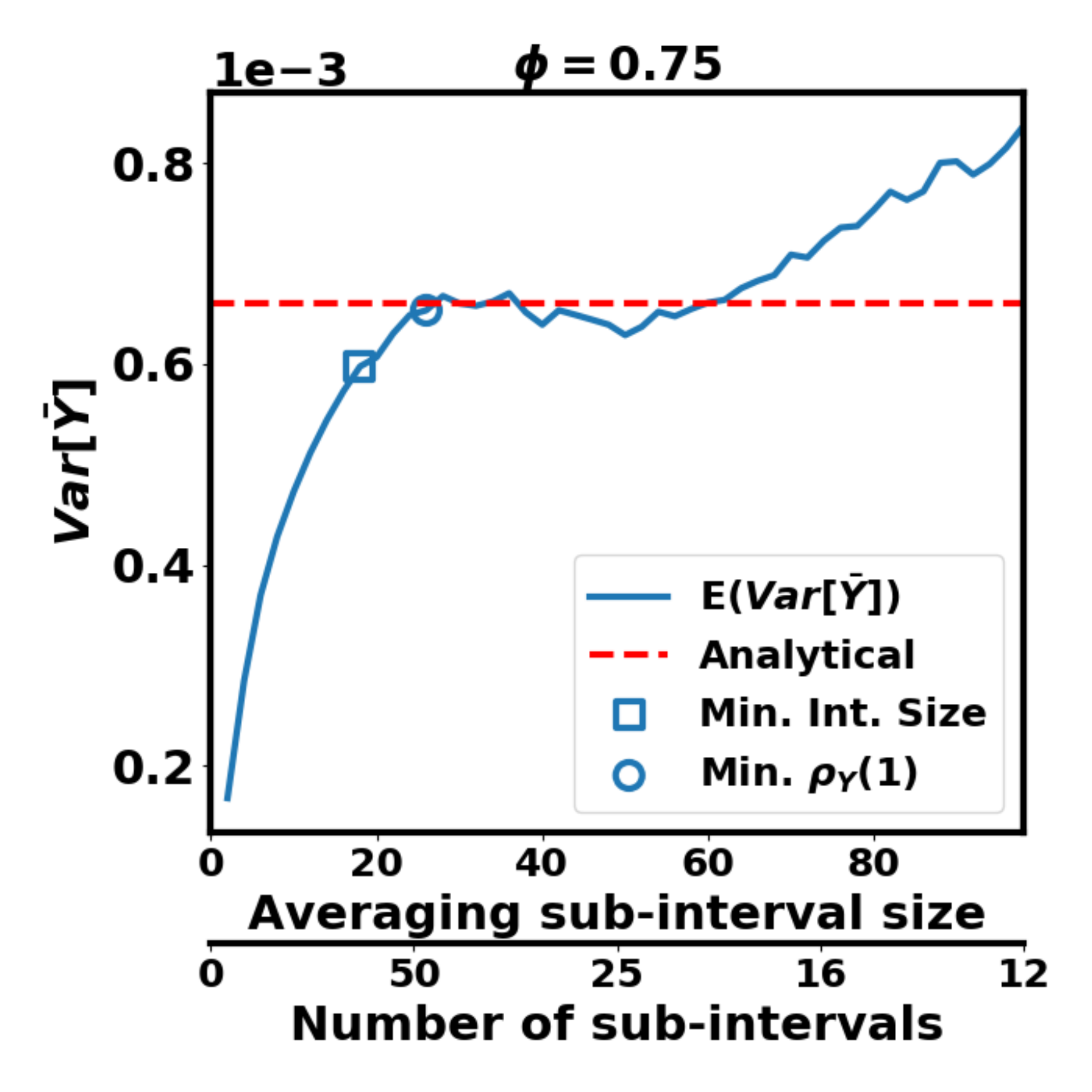}
  \end{minipage}
  \begin{minipage}[b]{0.325\textwidth}
    \includegraphics[trim={0cm 0 0 0},clip, width=1.\textwidth]{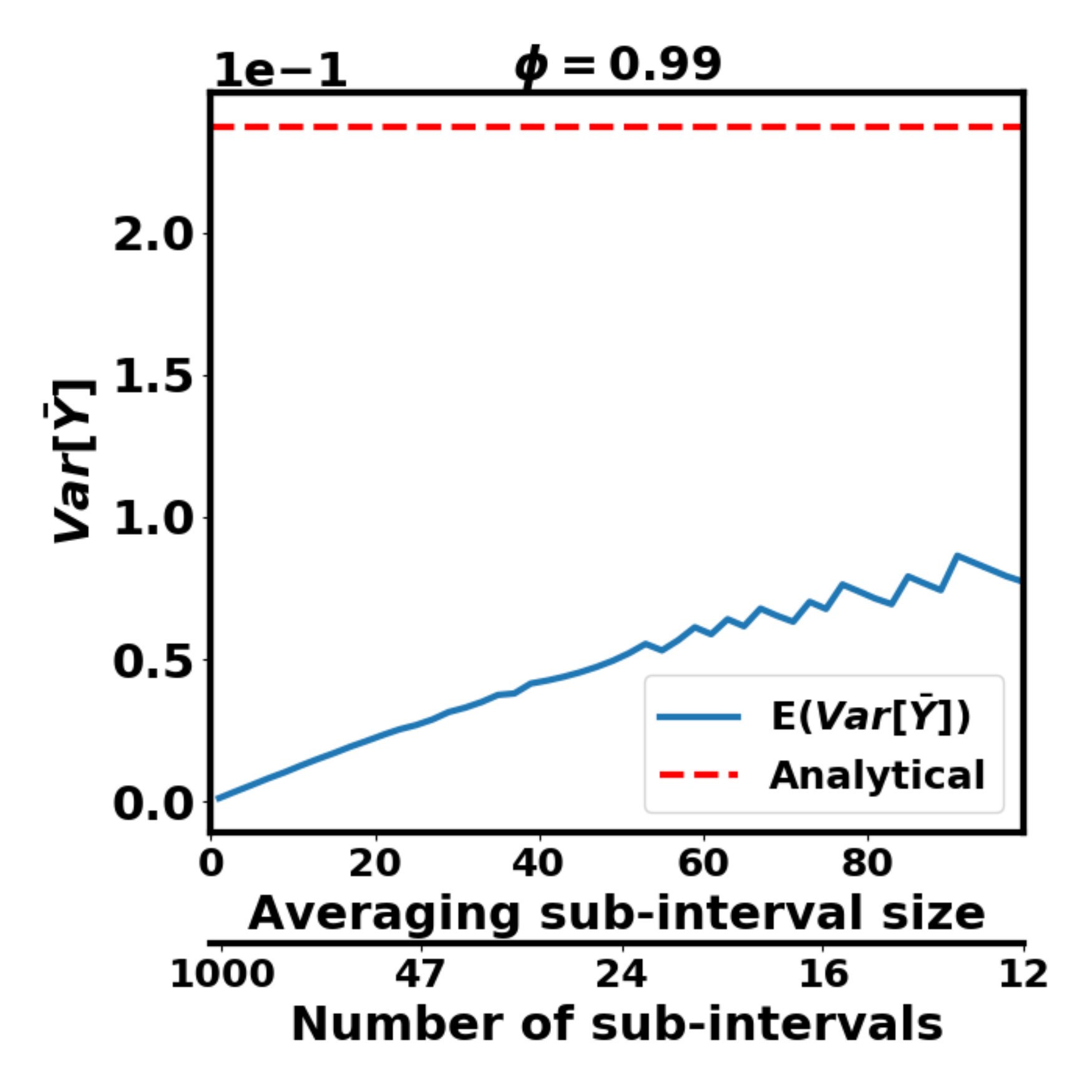}
  \end{minipage}
  \caption{Mean variance of sub-interval averages, $Var[\bar{Y}]$, of ARMA$(1,0)$ process as a function of sub-window size with 1000 samples for different values of $\phi_1 = \{0, 0.75, 0.99\}$ compared with the analytical value of variance.}
  \label{fig:subwinvar}
\end{figure*}

Since the lag autocorrelation of sub-interval averages diminish for larger lags in a stochastic process, we only focus on the uncorrelated first lags for determining the minimum sub-interval size. We also show the changes in the variance of the mean distribution, in Fig. \ref{fig:subwinvar}. We observe in autocorrelated cases if the sample size is large enough (such as the case in Fig. \ref{fig:armarho1}b and Fig. \ref{fig:subwinvar}b), at a certain sub-interval size, $\rho_Y(1)$ is less than the standard error of sample size. Traditionally, the sub-interval size is determined as the minimum interval size needed to calculate the variance of series. However, when comparing against the analytical solution, we observe this estimate of uncertainty still exhibits some error in comparison with the analytical value. Since we observe that minimizing the error is obtained when $\rho_y(1) \simeq 0$, we propose that a better approach for this technique is to use the largest number of sub-windows possible which meets this condition. Ideally, where the autocorrelation of all the sub-interval lags are zero, the variance estimate is exact. However, here we focused on selecting the sub-interval size that minimizes the first lag value for getting a simple yet relatively accurate variance estimate.

In Table \ref{tbl:compmethods} we compare the variance measurement techniques against the analytical values. When $\phi_1 = 0$, the signal is white noise and completely uncorrelated, thus the analytical value of the mean variance becomes equivalent to the variance of sampling distribution of the mean obtained from the Lindberg-Levy Central Limit Theorem for independent data. Moreover, in this case, the best window length for sub-interval averaging is the same timestep as the original series, as shown in Figs. \ref{fig:armarho1}(a) and \ref{fig:subwinvar}(a). On the other hand, using the integral correlation time approach introduces finite lag effects within the kernel estimator integration, hence the integral correlation time estimate of the mean variance exhibits small error.

\begin{table}[ht!]
\caption{\label{tbl:compmethods} Mean variance of ARMA$(1,0)$ using different methods for $n=1000$ samples.}
\setlength\tabcolsep{0pt}
\begin{ruledtabular}
\begin{tabularx}{0.95\textwidth}{ >{\centering\arraybackslash}m{1.2in} | >{\centering\arraybackslash}m{1.4in} | >{\centering\arraybackslash}m{1.4in} | >{\centering\arraybackslash}m{1.4in} | >{\centering\arraybackslash}m{1.4in}  }
 Method & Analytical & Treating as uncorrelated signals & Integral Correlation Time & Sub-interval Averaging \\
 \hline
  & ${\sigma_X^2 (1 + \phi_1) \over n (1 - \phi_1)}$ & ${\sigma_X^2 \over {n}}$ &  ${\tau_{int}}{\sigma_X^2 \over n}$ & ${\sigma_{{Y}}^2 \over {T}}$ \\
 \hline
 $\phi_1 = 0$ & $4.05 \times 10^{-5}$ & $4.05 \times 10^{-5}$ & $3.93\times 10^{-5}$ & $4.05 \times 10^{-5}$ \\
 \hline
 $\phi_1 = 0.75$ & $6.61 \times 10^{-4}$ & $9.44 \times 10^{-5}$ & $2.76 \times 10^{-4}$ & $6.46 \times 10^{-4}$ \\
 \hline
 $\phi_1 = 0.99$ & $0.12$ & $8.85 \times 10^{-4}$ & $0.037$ & $-$ \\
\end{tabularx}
\end{ruledtabular}
\end{table}

In the case of $\phi=0.75$, we observe due to finite autocorrelation of lags, the analytical value of mean variance is larger than treating the time series as uncorrelated. One should bear in mind that the mean variance is always larger than treating the time series as uncorrelated, and neglecting this fact can result in underestimation of temporal uncertainties. Nonetheless, with $n=1000$ samples we have a sufficient number of time-steps that the so that the sub-interval averaging mean variance asymptotes to the analytical value of the sample mean variance (see Fig. \ref{fig:subwinvar}b) and be used as a good estimate of the actual analytical mean variance in this case. Again we have chosen the largest number of sub-interval sub-windows ($T=55$) value where $\rho_Y(1)$ is insignificant within confidence intervals (see Fig. \ref{fig:armarho1}b). On the other hand, although the integral correlation time method does a better job than simply treating $X$ as an uncorrelated signal, it  suffers from the errors of autocorrelation lags due to finite sample size of time series. The integral correlation time method shows significant sensitivity to both the choice of kernel estimator and insignificance threshold of the autocorrelated lags. From this comparison, it is obvious that the sub-interval averaging is a better choice for estimating the mean variance.

\begin{figure*}[!htb]
  \centering
  \begin{minipage}[b]{0.37\textwidth}
    \includegraphics[trim={0cm 0 0 0},clip, width=1.\textwidth]{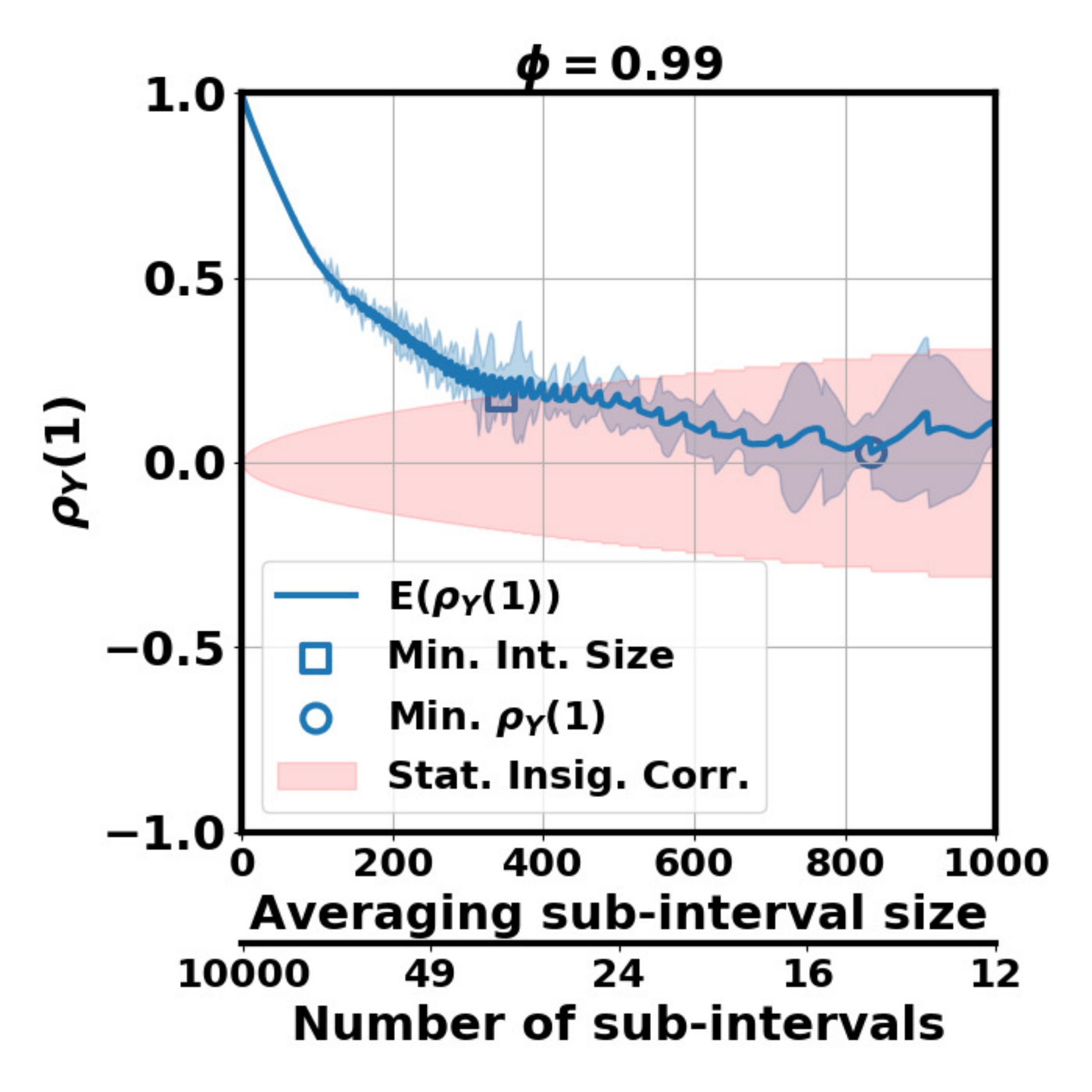}
  \end{minipage}
  \begin{minipage}[b]{0.37\textwidth}
    \includegraphics[trim={0cm 0 0 0},clip, width=1.\textwidth]{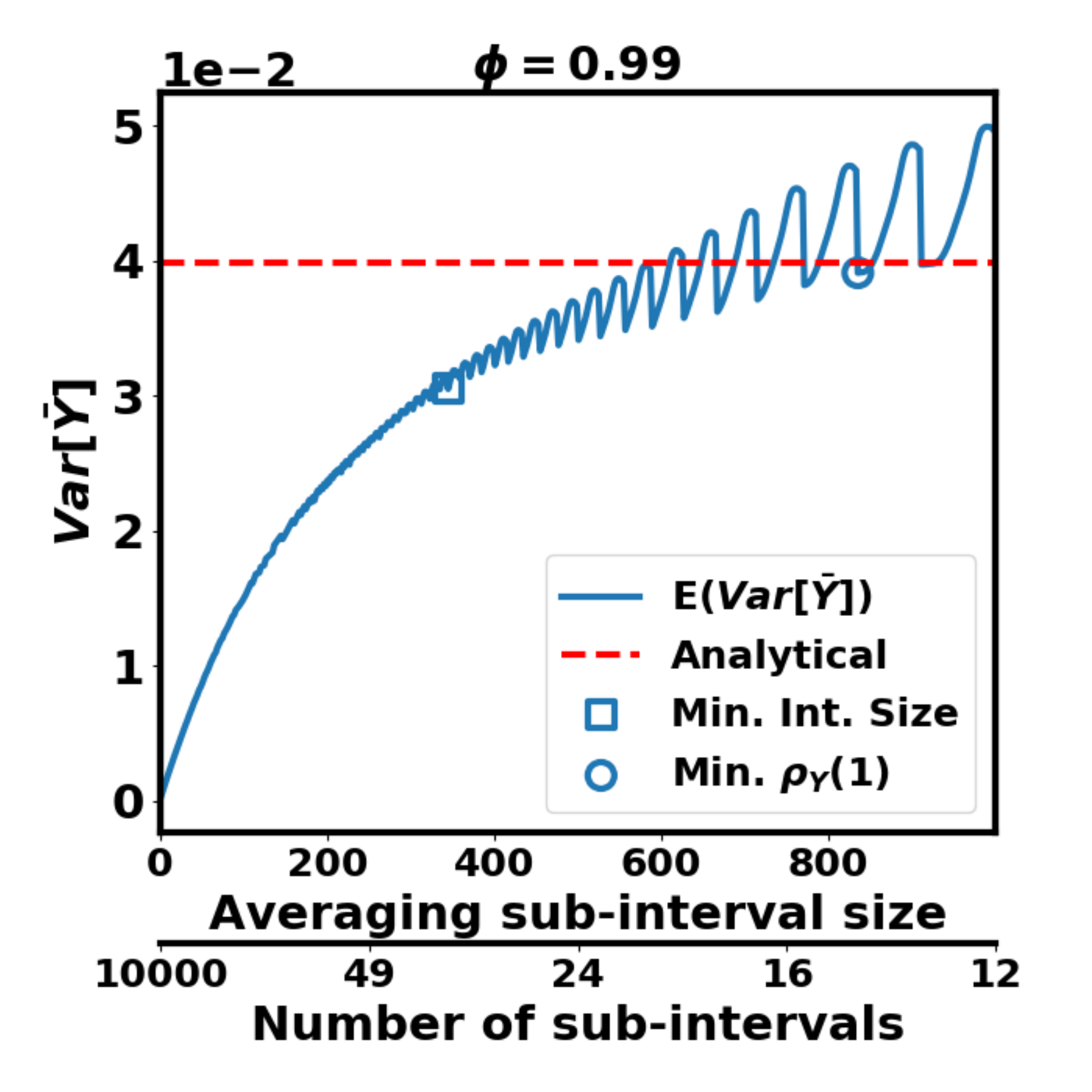}
  \end{minipage}
    \caption{(a) Autocorrelation of first sub-window average lag, $\rho_Y(1)$, and (b) Mean variance of sub-interval averages, $Var[\bar{Y}]$, of ARMA$(1,0)$ process as a function of sub-window size with 10000 samples for $\phi_1 = 0.99$ compared against the analytical value of variance.}
    \label{fig:largertssub}
\end{figure*}

In the case of $\phi_1 = 0.99$, we are dealing with a long memory process with many correlated lags, where the number of time-steps used to generate the signal is not enough for a good estimate of the mean variance with either one of our considered approaches. Using the integral correlation time method, the mean variance exhibits large error, mainly due to fact that the standard error of autocorrelation function is large due to small sample size yet large correlation (see Fig. \ref{fig:armaacf}c). On the other hand, in Fig. \ref{fig:subwinvar}(c) we can observe the mean variance of sub-intervals has not converged yet, and at no point does $\rho_y(1)$ enter the statistical insignificance region in Fig. \ref{fig:armarho1}(c). Such a case can happen in plasma turbulence simulations with very slow dynamics, in these cases the simulation needs to be run longer for an accurate assessment of the mean variance. We note that in the $\phi_1=0.99$ case, by merely having larger time series, the sub-interval mean variance technique converges to the analytical variance value, as shown in Fig. \ref{fig:largertssub}. 

\section{Assessing Mean and Mean Variance Convergence of Simulated Turbulence Quantities} \label{simulationvar}

From Sec. \ref{sec:armatest}, we observed that $Var[\bar{Y}]$ estimate of sub-interval averages technique is a reasonable approach for estimating the variance of the mean distribution as long as we have a simulation time length much larger than autocorrelation timescale. One could run the simulation long enough and perform the sub-interval averaging of measurements to check if the mean variance is small compare to the mean quantity, and if the mean of turbulence quantity can be confidently used for validation or calibration purposes. However, if the simulation is not of sufficient length to confidently estimate the variance, these approaches offer no guidance on how much more simulation time is needed to accurately estimate the mean variance.

Nevertheless, modeling the stochastic process of turbulence quantity time series within saturation phase of simulation can provide us with the means of forecasting the temporal uncertainty at later simulation times. Hence, in this Section, we study the fitting of Gaussian ARMA processes to the gyrokinetic simulation energy flux within the saturation phase in order to determine the process of turbulence quantity. By determining which process best fits the data, we can estimate and forecast the mean variance of a turbulence quantity at later simulation times. Specifically, we use ARMA models to determine the process of the turbulence quantity and assess if a simulation has been run long enough. Once we determine the process of the simulation, we can use the fitted ARMA model coefficients to forecast how the variance of mean decreases for later simulation times. This study builds upon recent work by Parker \textit{et. al.}\cite{parker2018}, examining the performance of ARMA extrapolation methods for both strongly and weakly driven turbulence.

\subsection{Simulations Detail}

For this study, we utilize gyrokinetic simulation predictions of the ion energy flux $Q_i$, using parameters taken from a series of ion-temperature gradient (ITG) mode dominated neutral beam heated DIII-D tokamak high-confinement mode (H-mode) plasmas, the details of which can be found in Luce \textit{et al.}\cite{luce2017}. More specifically, we utilize the results of three different simulations corresponding to three different values of the local normalized ion temperature gradient inverse scale length $a/L_{Ti} = -a d ln(T_i)/dr$, where $r$ is the minor radius of a flux surface at the outboard midplane and $a$ its value at the separatrix. The first simulation uses parameters corresponding to the nominal measured value of $a/L_{Ti}$ (as determined by standard profile curve-fitting analysis) at $\rho_{tor} = 0.6$ in a discharge with approximately $7$ MW of injected heating power but only $1.4$ N-m of injected torque, while the other two simulations use values of $a/L_{T_i}$ equal to $80\%$ and $50\%$ of the measured value, respectively.  All other input parameters are held fixed at their measured values, which allows us to systematically quantify how the turbulence temporal characteristics change as the ITG mode drive is reduced.

The simulations were performed with the nonlinear initial value continuum gyrokinetic code CGYRO\cite{candy2016}. The simulations span a domain size of $111\rho_s$ by $63\rho_s$ in the radial and binormal directions, where $\rho_s = c_s/\Omega_{c_i}$ is the ion sound-speed gyroradius. The simulations are fully spectral in the perpendicular plane, and include 320 radial modenumbers (resolving up to a maximum $k_x \rho_s = 9.0$) and 12 binormal modes (spanning $0.1 \le k_y \rho_s \le 1.1$).  Parallel motion derivatives are treated with a sixth-order conservative upwind finite differencing scheme using 24 grid points in $\theta$. Velocity space is represented using the same ($\xi$,$v$) coordinates as the neoclassical NEO code\cite{belli2008,belli2012}, where $\xi = v_\parallel/v$ is the cosine of the pitch angle and $v$ the speed; 24 grid points in $\xi$ and 8 in $v$ are used. The simulations are local, and include magnetic flux surface shaping through the Miller representation\cite{miller1998,candy2009}, transverse magnetic fluctuations, electron and ion collisions, and equilibrium rotation and shear effects treated with a novel wavenumber advection algorithm\cite{candy2018}.  Three ion species are included- thermal deuterium and carbon, as well as fast beam ions (modeled as having Maxwellian distribution with $T_{fast}/T_e = 12.4$, whereas for thermal ions $T_i/T_e = 1.16$), but only transport from the thermal ions is considered here. All particle species (ions and electrons) are treated fully gyrokinetically.

\begin{figure*}[!htb]
  \centering
  \begin{minipage}[b]{0.325\textwidth}
    \includegraphics[trim={1.75cm 0.5cm 1cm 0},clip, width=1.\textwidth]{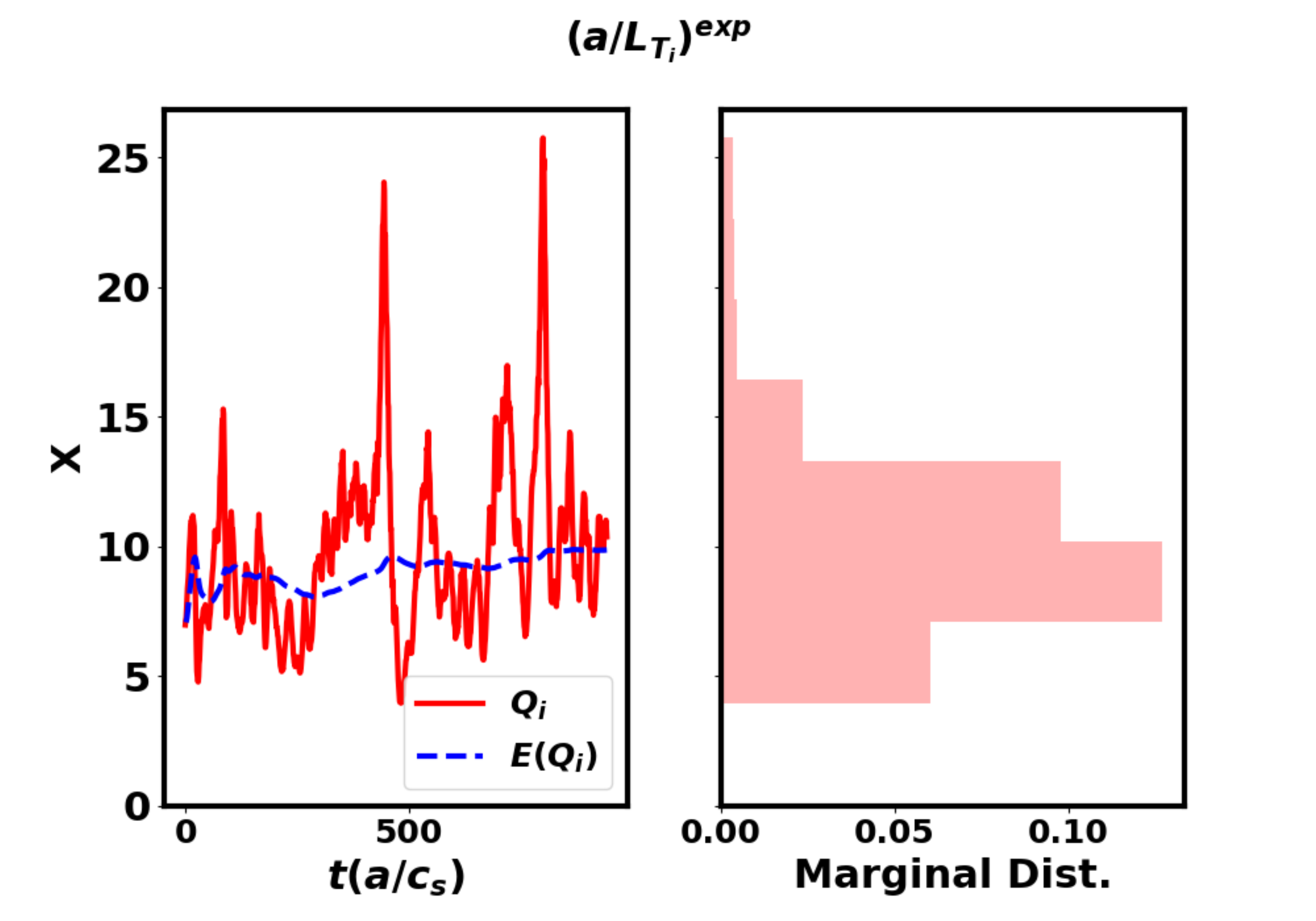}
  \end{minipage}
  \begin{minipage}[b]{0.325\textwidth}
    \includegraphics[trim={1.75cm 0.5cm 1cm 0},clip, width=1.\textwidth]{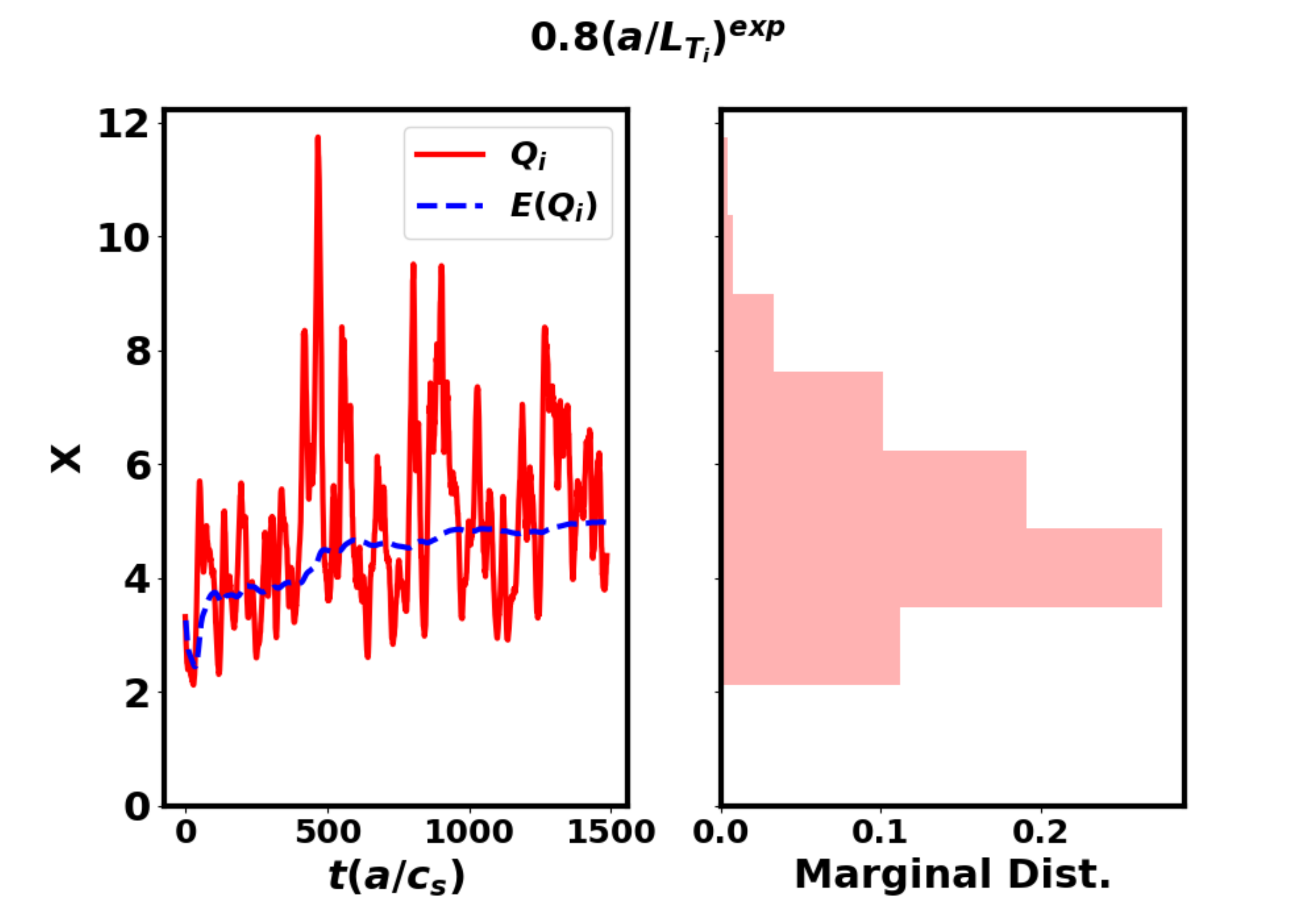}
  \end{minipage}
  \begin{minipage}[b]{0.325\textwidth}
    \includegraphics[trim={.5cm 0.5cm 2.25cm 0},clip, width=1.\textwidth]{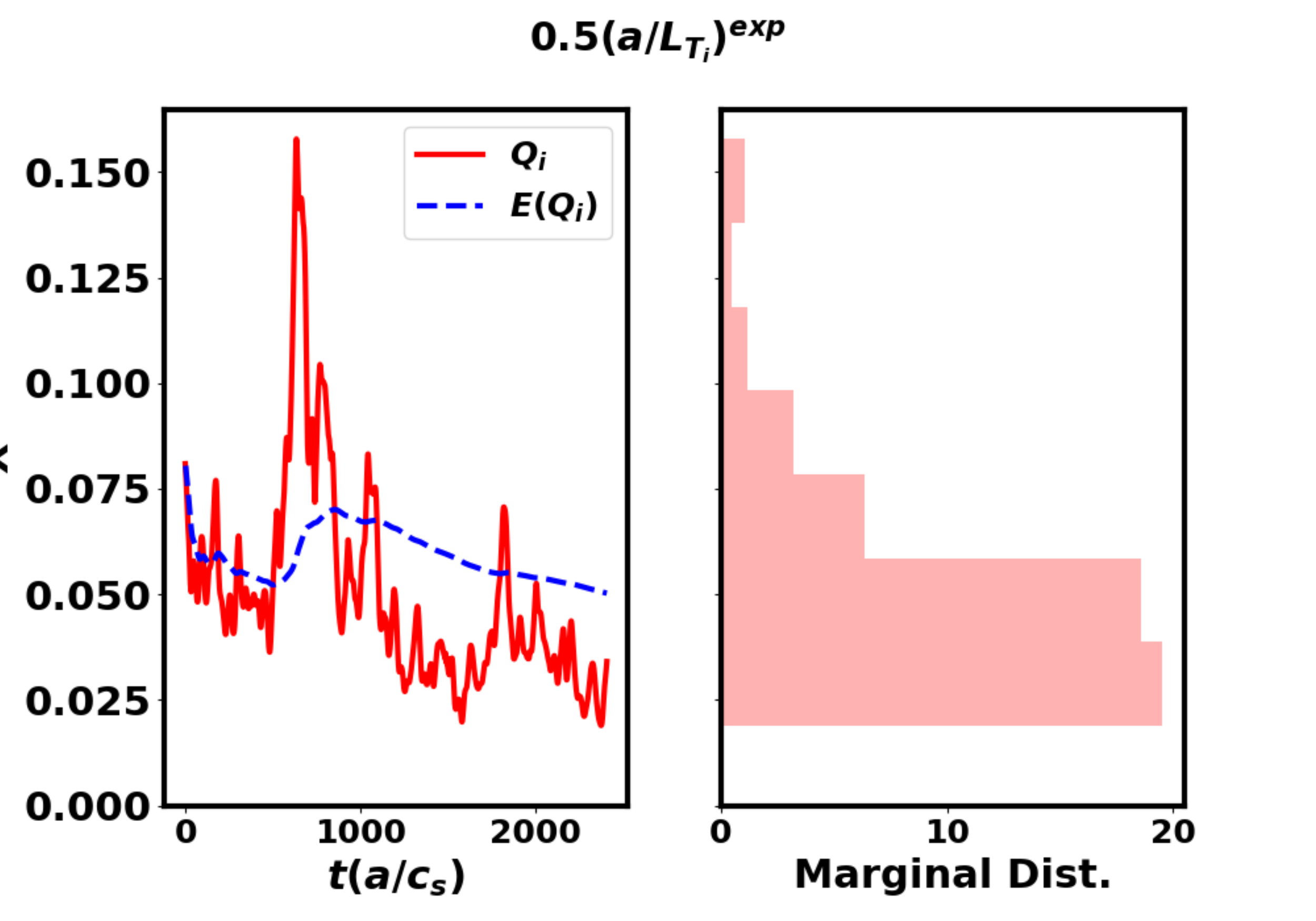}
  \end{minipage}
  \caption{Ion energy flux time series with their cumulative mean, and the marginal distribution of ion energy flux gyrokenitic simulations of ion temperature length scale: (a) $a/L_{T_i} = (a/L_{T_i})^{exp}$, (b) $a/L_{T_i} = 0.8(a/L_{T_i})^{exp}$, (c) $a/L_{T_i} = 0.5(a/L_{T_i})^{exp}$. The histograms show simulation temporal sample distribution. Simulation times before saturation phase are not shown. Fluxes are normalized to gyro-Bohm energy flux.}
  \label{fig:simts}
\end{figure*}

\begin{figure*}[!htb]
  \centering
  \begin{minipage}[b]{0.325\textwidth}
    \includegraphics[trim={.25cm 0.75cm 2cm 0},clip, width=1.\textwidth]{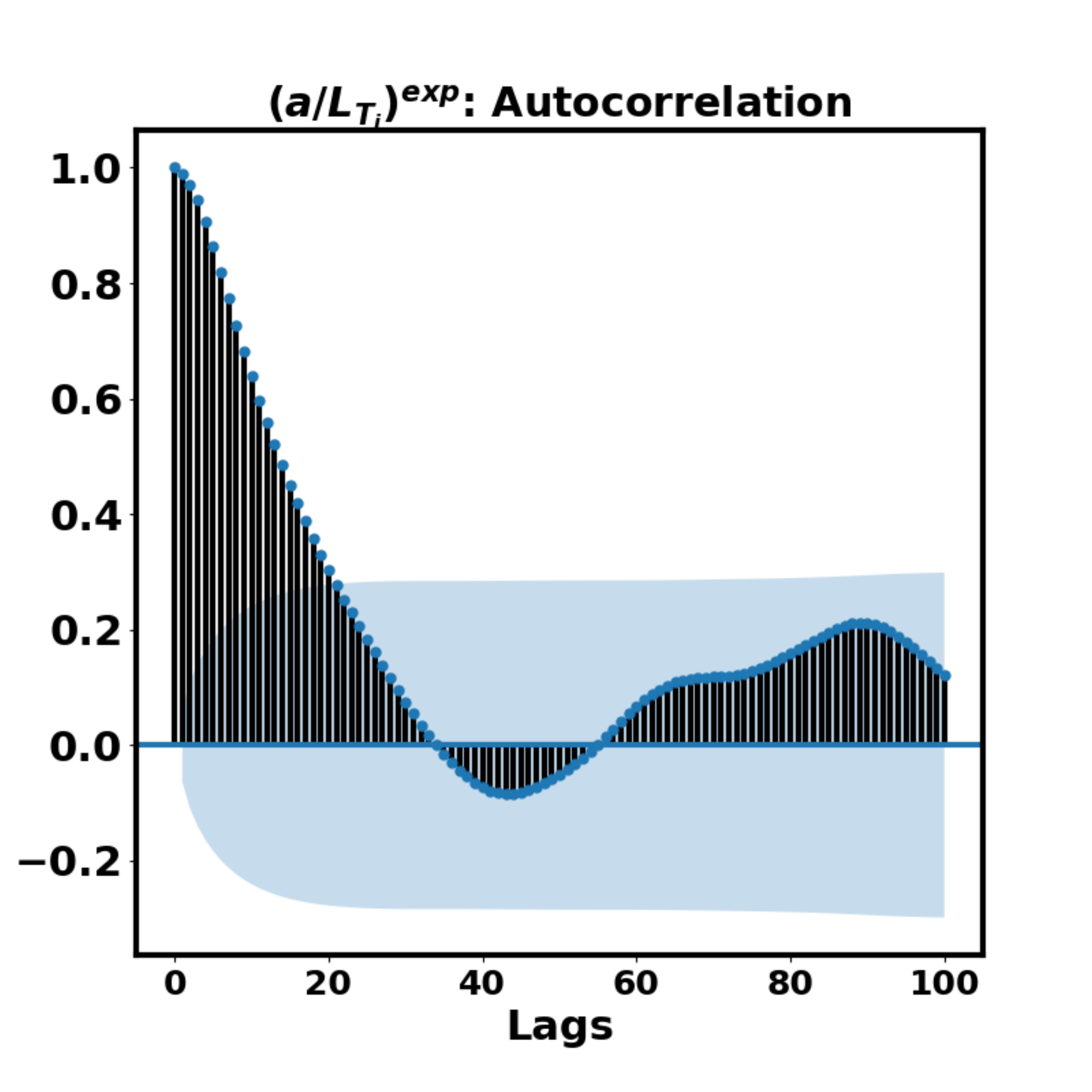}
  \end{minipage}
  \begin{minipage}[b]{0.325\textwidth}
    \includegraphics[trim={.25cm 0.75cm 2cm 0},clip, width=1.\textwidth]{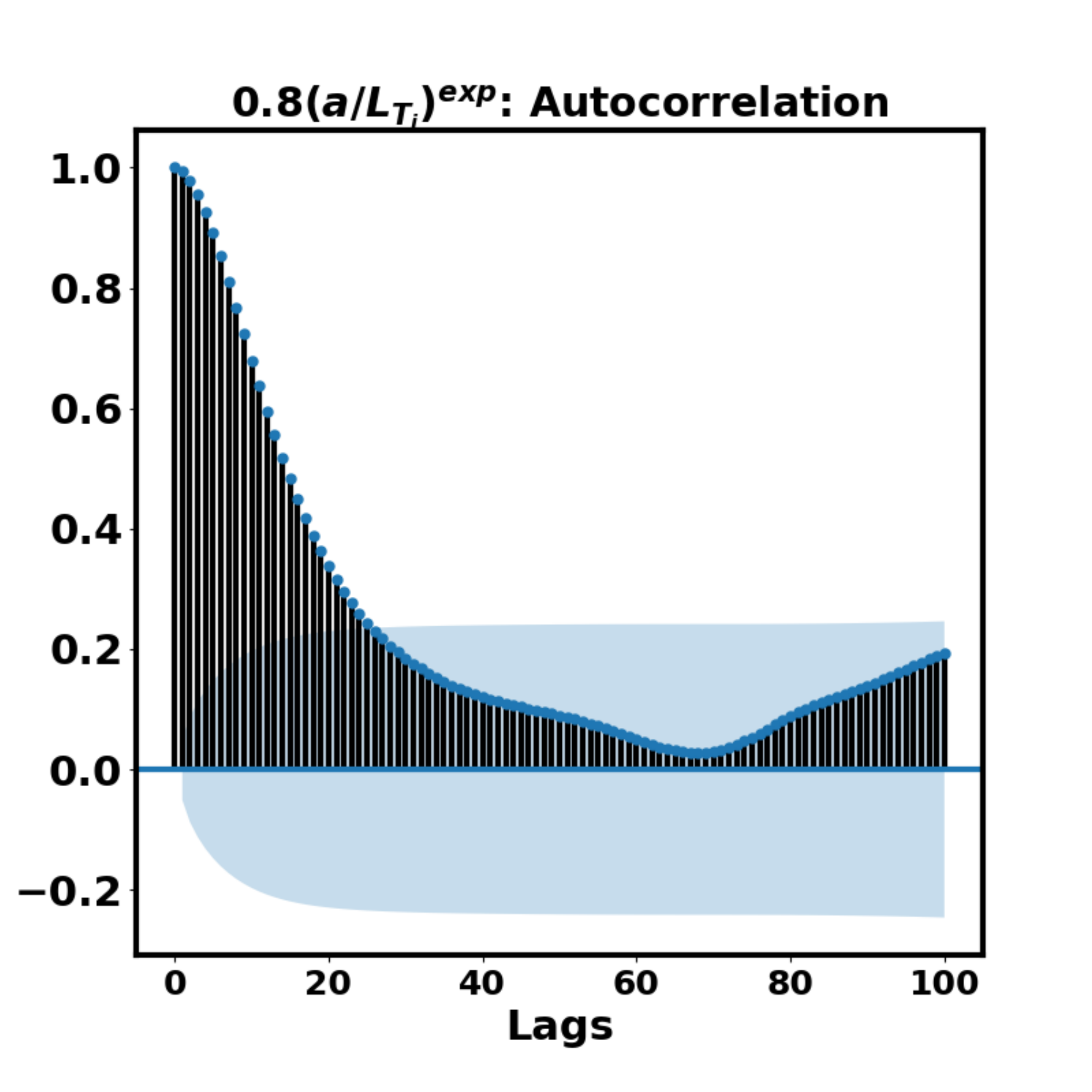}
  \end{minipage}
  \begin{minipage}[b]{0.325\textwidth}
    \includegraphics[trim={.25cm 0.75cm 2cm 0},clip, width=1.\textwidth]{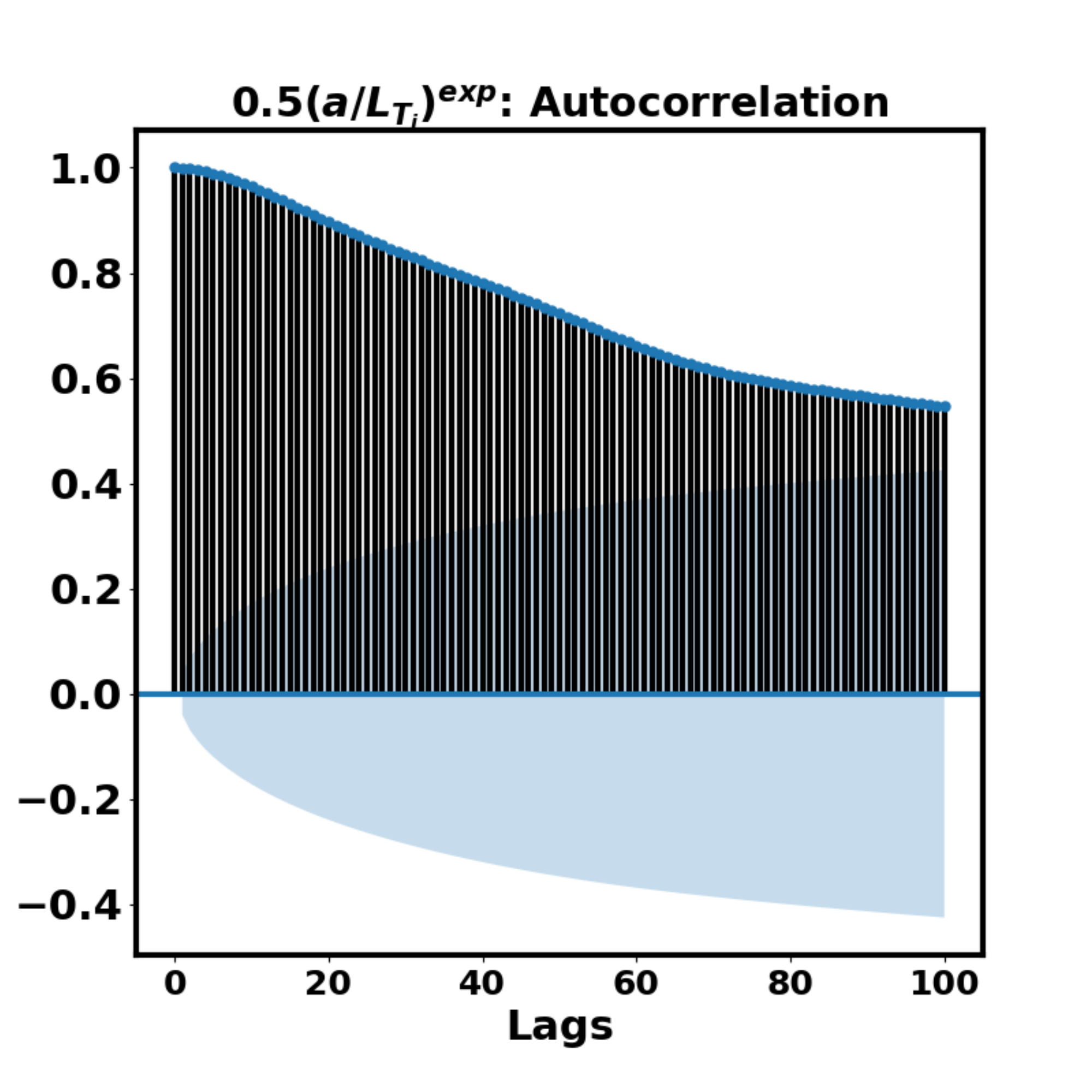}
  \end{minipage}
  \caption{Autocorrelation function of ion energy flux gyrokenitic simulations of ion temperature length scale: (a) $a/L_{T_i} = (a/L_{T_i})^{exp}$, (b) $a/L_{T_i} = 0.8(a/L_{T_i})^{exp}$, (c) $a/L_{T_i} = 0.5(a/L_{T_i})^{exp}$. The shaded area shows statistically insignificant autocorrelation within 95 percent confidence interval.}
  \label{fig:simacf}
\end{figure*}

The time series of energy fluxes for $1000 (a/c_s)$ or longer with a sampling rate of $1 (a/c_s)$ are shown in Fig. \ref{fig:simts} along with their running mean values through out the simulation. Here, $c_s=\sqrt{T_e/m_i}$ is the local ion sound speed. We observe for the experimental value of $a/L_{T_i}$ which is well above the critical value for instability, the ion energy flux exhibits a near-normal distribution, while at lower gradients the time series distribution skewness increases. Examination of the running means for each case (plotted as dashed lines in Fig. \ref{fig:simts}) shows that for the $(a/L_{T_i})^{exp}$ case, the mean of ion energy flux converges after approximately $500 (a/c_s)$, in $0.8(a/L_{T_i})^{exp}$ the ion energy flux mean converges after about $1000(a/c_s)$, while for the near marginal $0.5(a/L_{T_i})^{exp}$ case, the mean still has not converged after $2500 (a/c_s)$. {Here, we have deliberately kept the simulation length short to analyze the convergence of the introduced methods in the non-converged case}. Consistent with the results, the $0.5(a/L_{T_i})^{exp}$ case autocorrelation function exhibits much a larger number of autocorrelated lags, as shown in Fig. \ref{fig:simacf}.

\subsection{Determining the Stochastic Process of Turbulence Quantity}

To model the stochastic process of a turbulence quantity, we propose to use the Box-Jenkins methodology\cite{box2015time} to fit a suitable order of ARMA model parameters to the simulation data, and obtain an approximate analytical relation that best fits the simulation quantity. {We note that the ARMA model assumes stationary time series and can only be used to fit the simulation time series after nonlinear saturation, where initial linear physics effects have completely vanished and there is good convergence in the mean of the turbulence quantity. Further discussion of determining the saturation phase start time and convergence of the mean can be be found in Ref.} \cite{holland2016}. After finding the best ARMA fit for the nonlinear phase, the long run variance can be determined and thereby the mean variance can easily and cost-effectively be calculated for a variety of the simulations lengths. From this information, the necessary simulation length can be determined.

To address ARMA fitting of skewed time series (e.g. near marginal case), Box and Cox\cite{boxcox1992} suggested use of an invertible link function $g(X_t)$ which transforms the original time series to a linear process where the ARMA fitting is appropriate. A popular link function that have been used in the literature for positively skewed marginal distributions, is the logarithm function $g(X_t) = \log{X_t}$. Different variations of generalized non-normal ARMA models have also been formulated\cite{benjamin2003,zheng2015}. For the simulations considered here we find logarithmic link function is sufficient, and defer the investigation of non-Gaussian processes to future studies.

Different orders of ARMA$(p,q)$ models can be fitted to $g(X_t)$ and the best $(p,q)$ can be selected based on standard statistical tests, e.g. significance of fit coefficients\cite{hannan1982}, insignificance autocorrelation of residual lags\cite{parzen1963}, normality of the residuals distribution\cite{dagostino1973}, and significance of lags Ljung-Box p-values\cite{ljung1978}. To assess the fitting statistics of ARMA series, here we considered $p$ and $q$ values ranging from zero to ten, and for statistical tests we considered the statistics of first ten lags against the $95\%$ confidence interval for the autocorrelation function, with an assumed p-value threshold of 0.05 for the null hypotheses. We again emphasize that ARMA models assume stationarity of the time series, therefore we can only applied where the evolving mean has almost converged. As a result, based on the evolving mean shown in Fig. \ref{fig:simts}, we applied ARMA fitting after $250 (a/c_s)$ for $(a/L_{T_i})^{exp}$ case, after $750 (a/c_s)$ to $0.8(a/L_{T_i})^{exp}$ case, and after $1750(a/c_s)$ to $0.5(a/L_{T_i})^{exp}$ case (which is not rigorously justifiable).

\begin{figure*}[!htb]
  \centering
  \begin{minipage}[b]{0.325\textwidth}
    \includegraphics[trim={.25cm 0.75cm 1cm 0},clip, width=1.\textwidth]{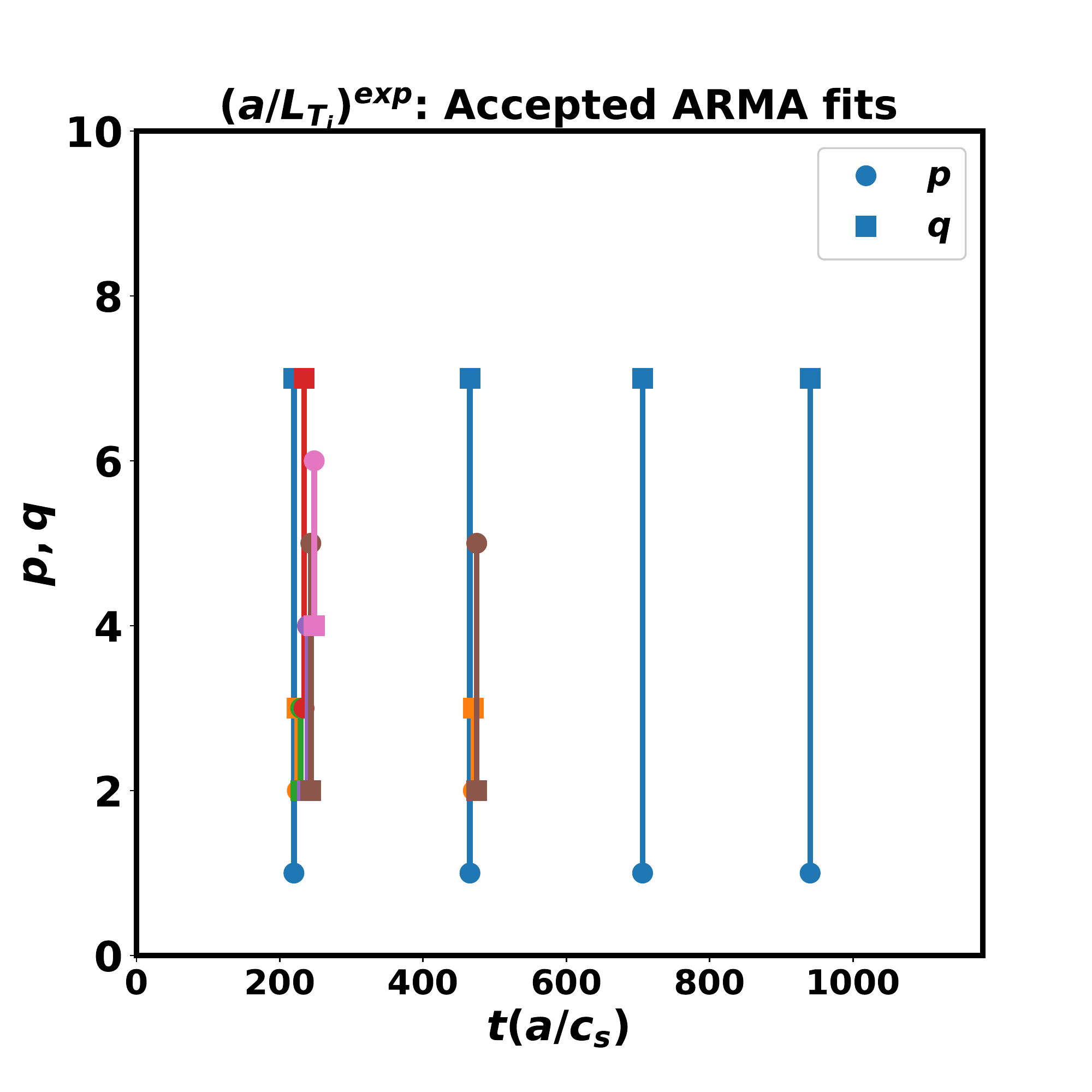}
  \end{minipage}
  \begin{minipage}[b]{0.325\textwidth}
    \includegraphics[trim={.25cm 0.75cm 1cm 0},clip, width=1.\textwidth]{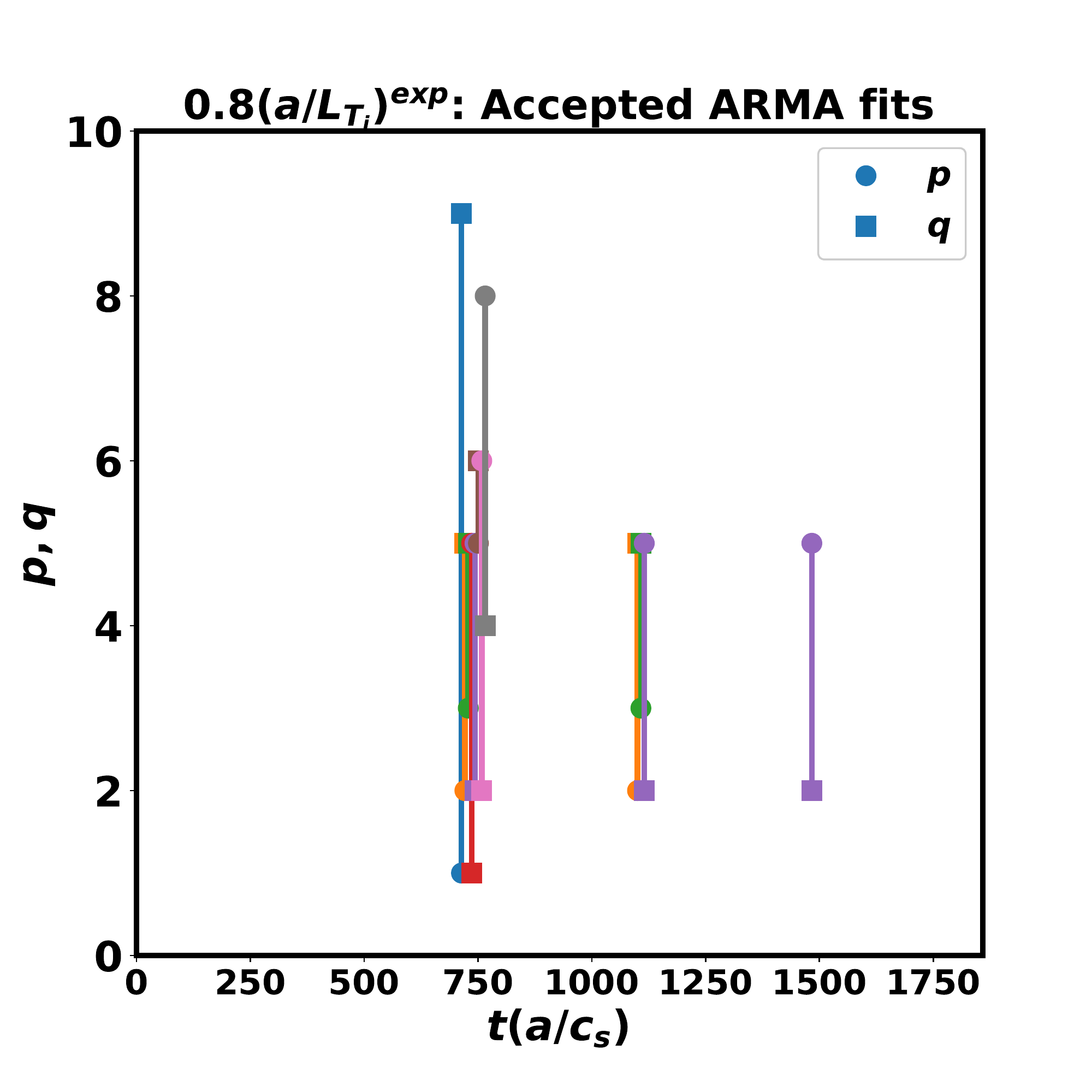}
  \end{minipage}
  \begin{minipage}[b]{0.325\textwidth}
    \includegraphics[trim={.25cm 0.75cm 1cm 0},clip, width=1.\textwidth]{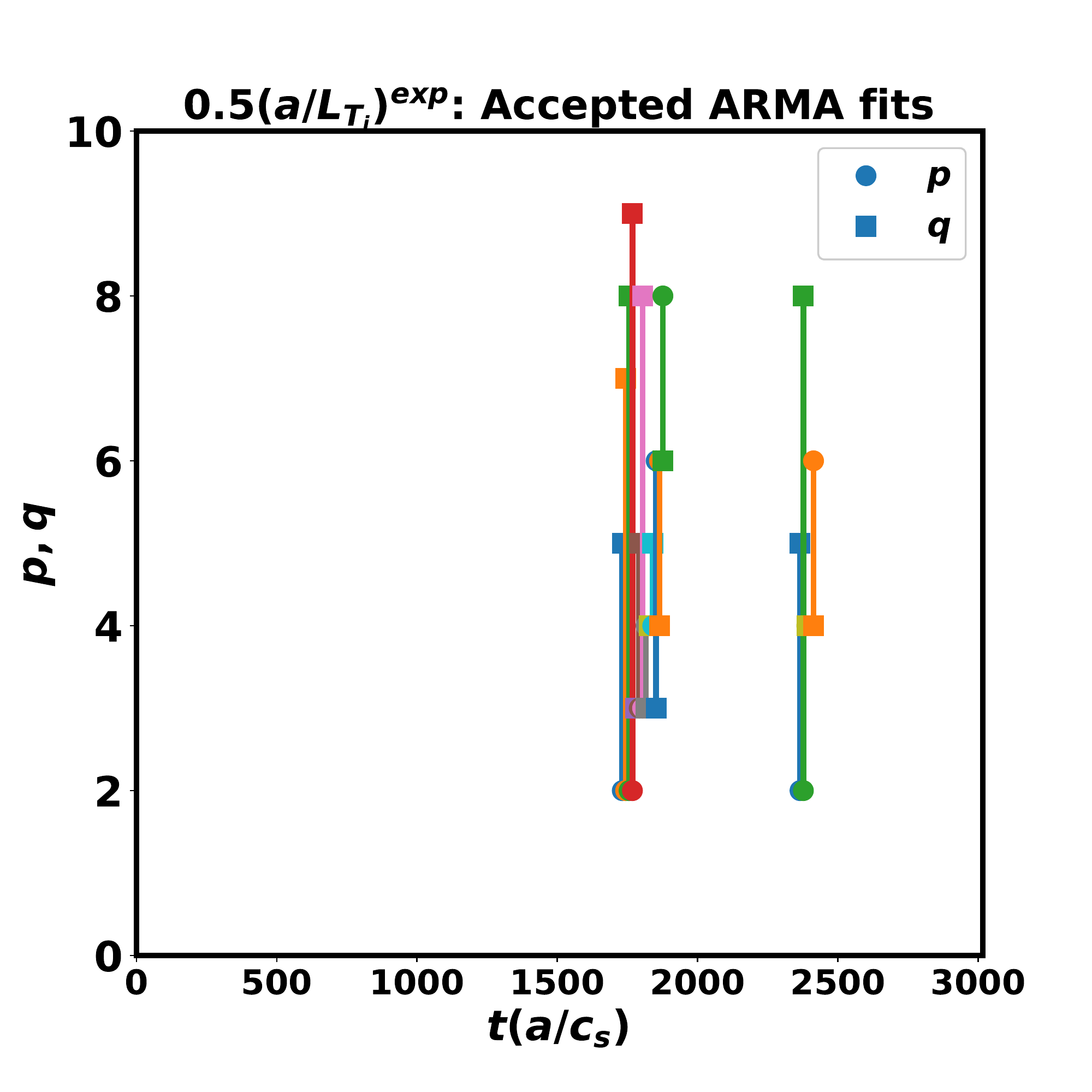}
  \end{minipage}
  \caption{Statistically feasible ARMA fits to the simulation data with respect to simulation timesteps for simulated ion energy flux: (a) $a/L_{T_i} = (a/L_{T_i})^{exp}$, (b) $a/L_{T_i} = 0.8(a/L_{T_i})^{exp}$, (c) $a/L_{T_i} = 0.5(a/L_{T_i})^{exp}$. {Distinct colors show different ARMA models.}}
  \label{fig:armafitscan}
\end{figure*}

In Fig. \ref{fig:armafitscan}, we have shown all ARMA fits that passed the statistical tests for different simulation cases, as a function of fitting window length. We observe at earlier simulation timesteps, the simulation data is not sufficient to accurately describe the stochastic process of turbulence. Hence, many different ARMA processes can fit the simulation data and pass the statistical tests. However, with the increase of simulation data points the stochasticity of the turbulence is better determined statistically. We observe after $750 (a/c_s)$ only one process ($ARMA(1,7)$) passes the tests for the $(a/L_{T_i})^{exp}$ simulation case. For the $0.8(a/L_{T_i})^{exp}$ simulation case, $ARMA(5,2)$ is the only process that can describe the simulation data after $1250 (a/c_s)$. However, in the near marginal $0.5(a/L_{T_i})^{exp}$ simulation case, there multiple feasible ARMA fits to the simulation data. This result can be associated to large autocorrelations and small sample sizes, and/or a break down in the assumption of Gaussian processes for the near marginal case. The reasoning behind such behavior can also be seen for autocorrelation function of ion energy flux for each simulation case (see Fig. \ref{fig:simacf}). Theoretically, the turbulence timescale is larger than current simulation length, showing that not all relevant timescale dynamics are resolved within the simulation. {We should note that the best practice in the case of getting multiple ARMA models for a time series is to run the simulation longer in order to rule out statistically infeasible ARMA models}\cite{ling2013uncertainty}{. If no ARMA model can fit a time series while the number of samples are very large, one can loosen the statistical hypotheses. However, if the number of samples are small while no ARMA fit can be found, we can deduce ARMA Gaussian process can not be applied to the simulation case, and non-Gaussian processes}\cite{benjamin2003} {should be modeled for that specific simulation case.}

\begin{figure*}[!htb]
  \centering
  \begin{minipage}[b]{0.325\textwidth}
    \includegraphics[trim={0cm 0cm 0cm 0}, width=.97\textwidth]{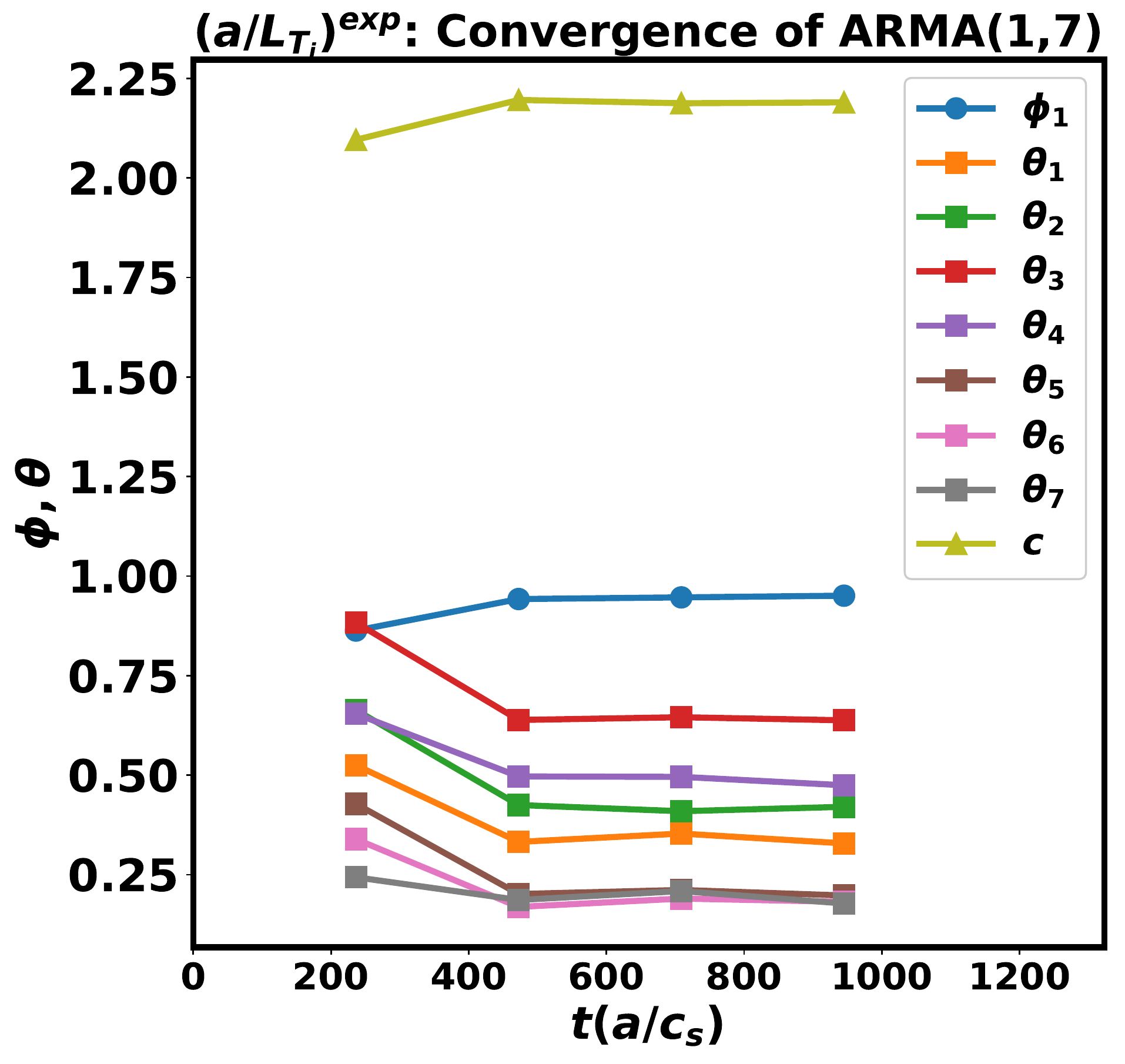}
  \end{minipage}
  \begin{minipage}[b]{0.325\textwidth}
    \includegraphics[trim={0cm 0cm 0cm 0}, width=1.\textwidth]{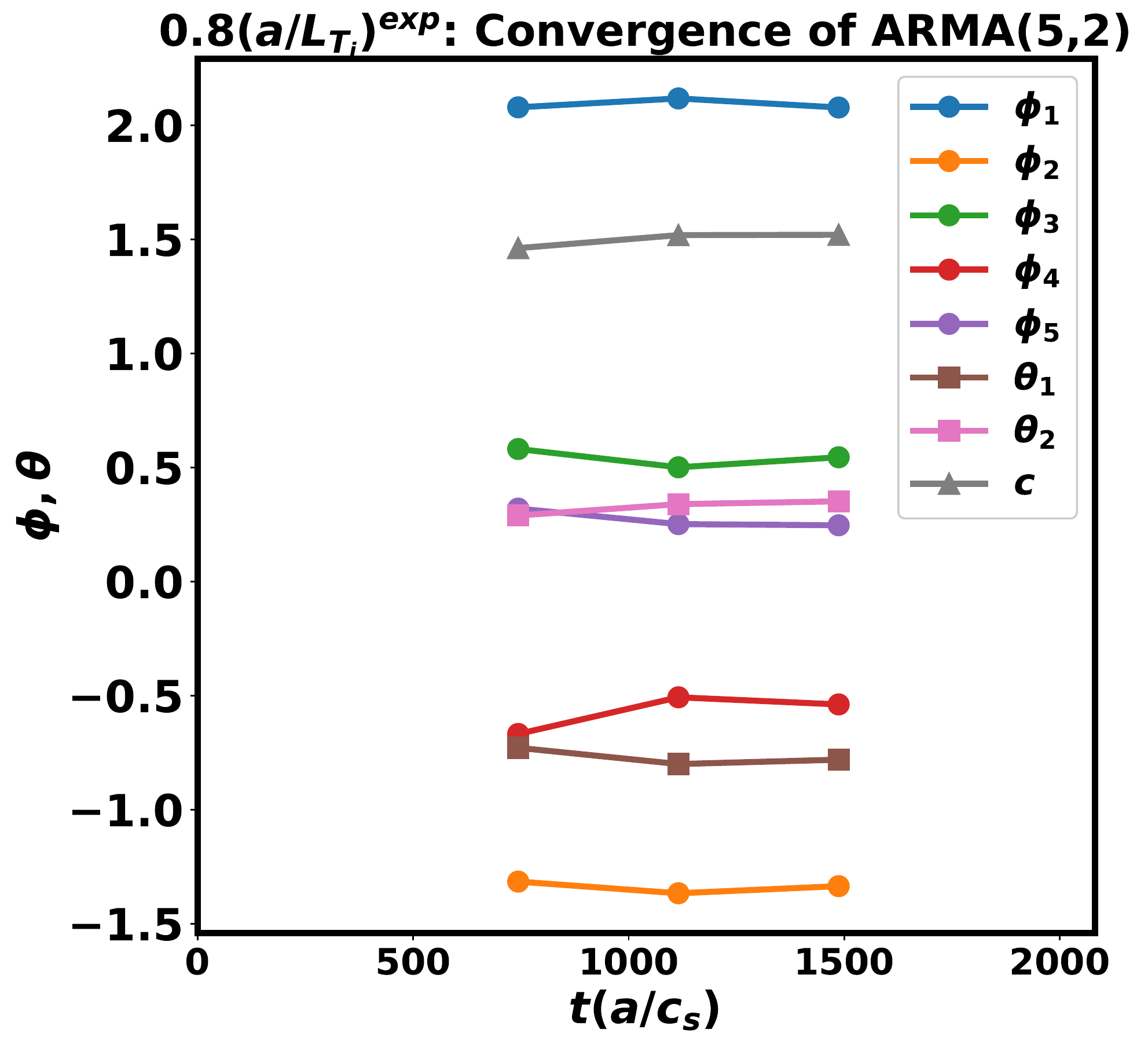}
  \end{minipage}
  \caption{Convergence of ARMA process coefficients for simulated ion energy flux for: (a) $a/L_{T_i} = (a/L_{T_i})^{exp}$, (b) $a/L_{T_i} = 0.8(a/L_{T_i})^{exp}$ as a function of simulation timestep. $ 0.8(a/L_{T_i})^{exp}$ case not shown since there is not a definite best ARMA model with current simulation length.}
  \label{fig:armacoefconv}
\end{figure*}

To further study the convergence of the ARMA model in the $(a/L_{T_i})^{exp}$ and $0.8(a/L_{T_i})^{exp}$ simulation cases, in Fig. \ref{fig:armacoefconv}, we have shown the changes in the ARMA constant offset (link-transformed time series mean), $\phi$ and $\theta$ coefficients. We observe after $500 (a/c_s)$ in the $(a/L_{T_i})^{exp}$ simulation case and after $1250 (a/c_s)$ in the $0.8(a/L_{T_i})^{exp}$ simulation case, the ARMA coefficients and the mean of turbulence quantity start to converge to a certain value, indicating there is enough simulation data to get a good estimate of mean and its fractional uncertainty.

\subsection{Forecasting the Mean Variance for Temporal Uncertainty Tolerance}

We extend our analysis to forecast the variance changes for later simulation times. These forecasts are performed by using the best ARMA model fits to generate additional data points for the $Q_i$ time series beyond the gyrokinetic results. Since the ARMA model inherently assumes the stationarity of the time series, one is able to forecast mean variance at later simulation time, once the stochastic process of turbulence is known. Here, we have used ARMA fitting of simulation data using a link function, extended the time series by continue simulating the time series of the based fitted ARMA, and calculate the mean variance of ARMA forecast using sub-interval averaging technique based on $\min(\rho_Y(1))$ criteria described in Sec. \ref{sec:armatest}.

\begin{figure*}[!htb]
  \centering
  \begin{minipage}[b]{0.325\textwidth}
    \includegraphics[trim={0cm 0cm 0cm 0},clip, width=1.\textwidth]{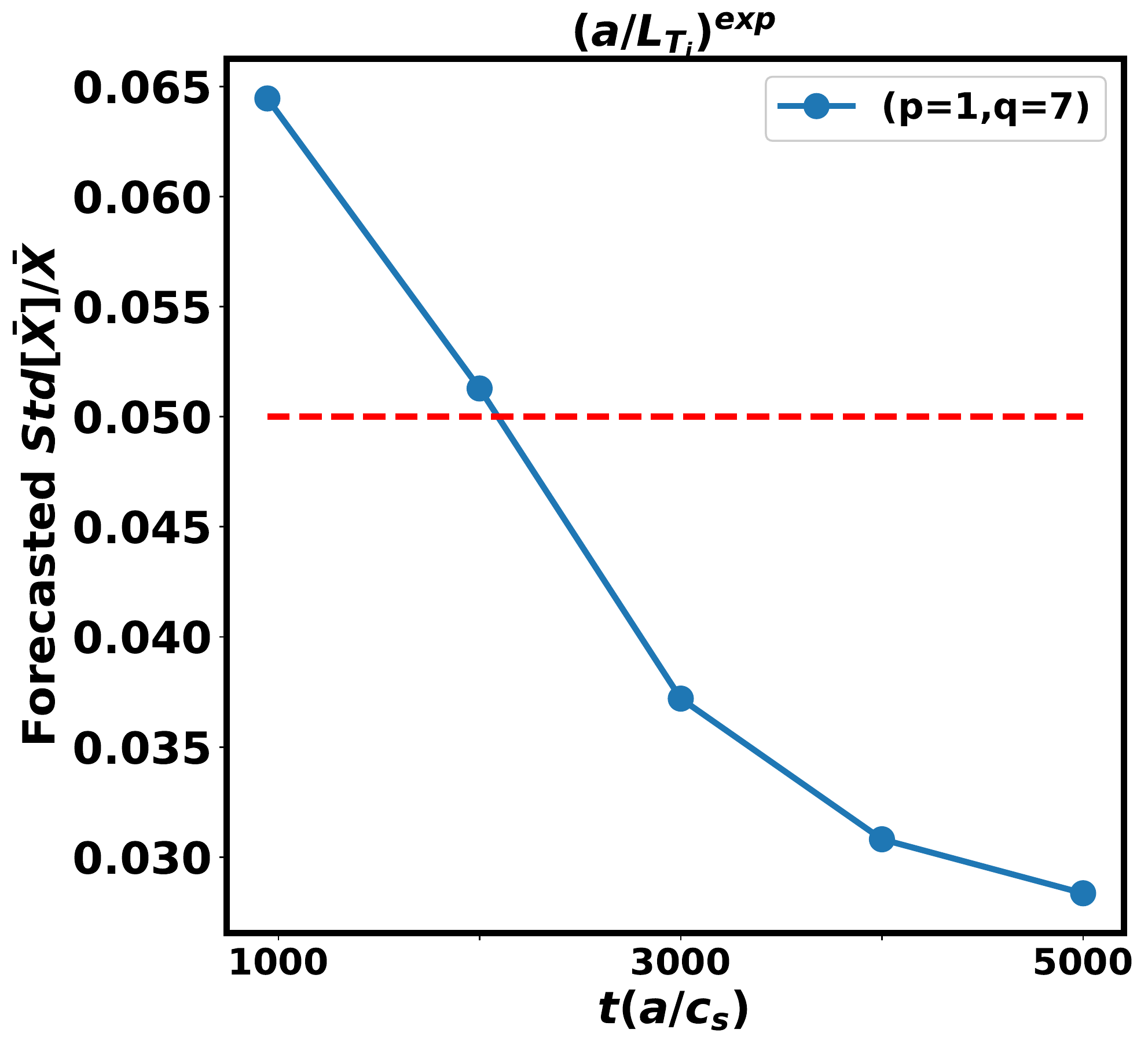}
  \end{minipage}
  \begin{minipage}[b]{0.325\textwidth}
    \includegraphics[trim={0cm 0cm 0cm 0},clip, width=1.\textwidth]{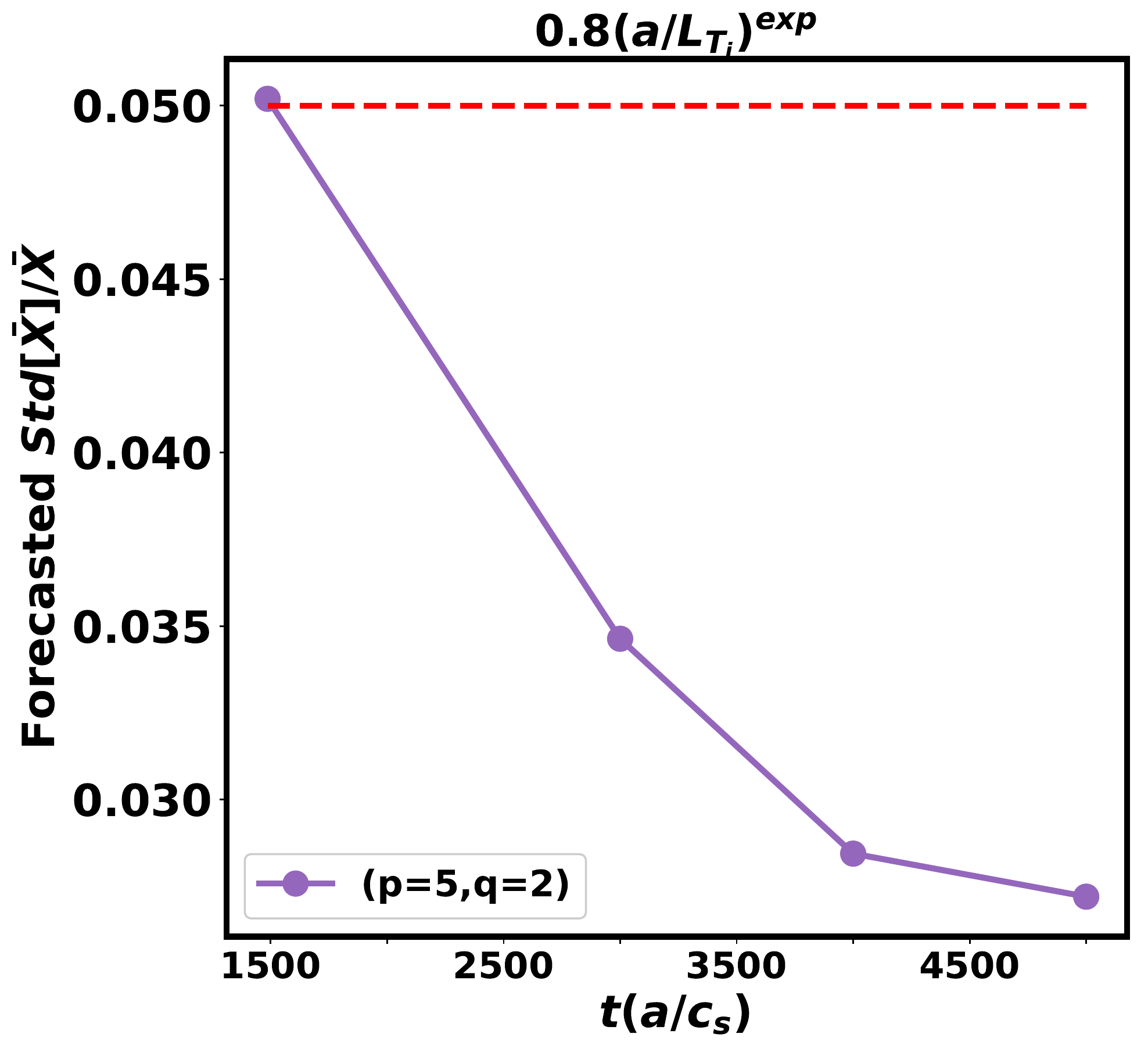}
  \end{minipage}
  \begin{minipage}[b]{0.325\textwidth}
    \includegraphics[trim={0cm 0cm 0cm 0},clip, width=1.\textwidth]{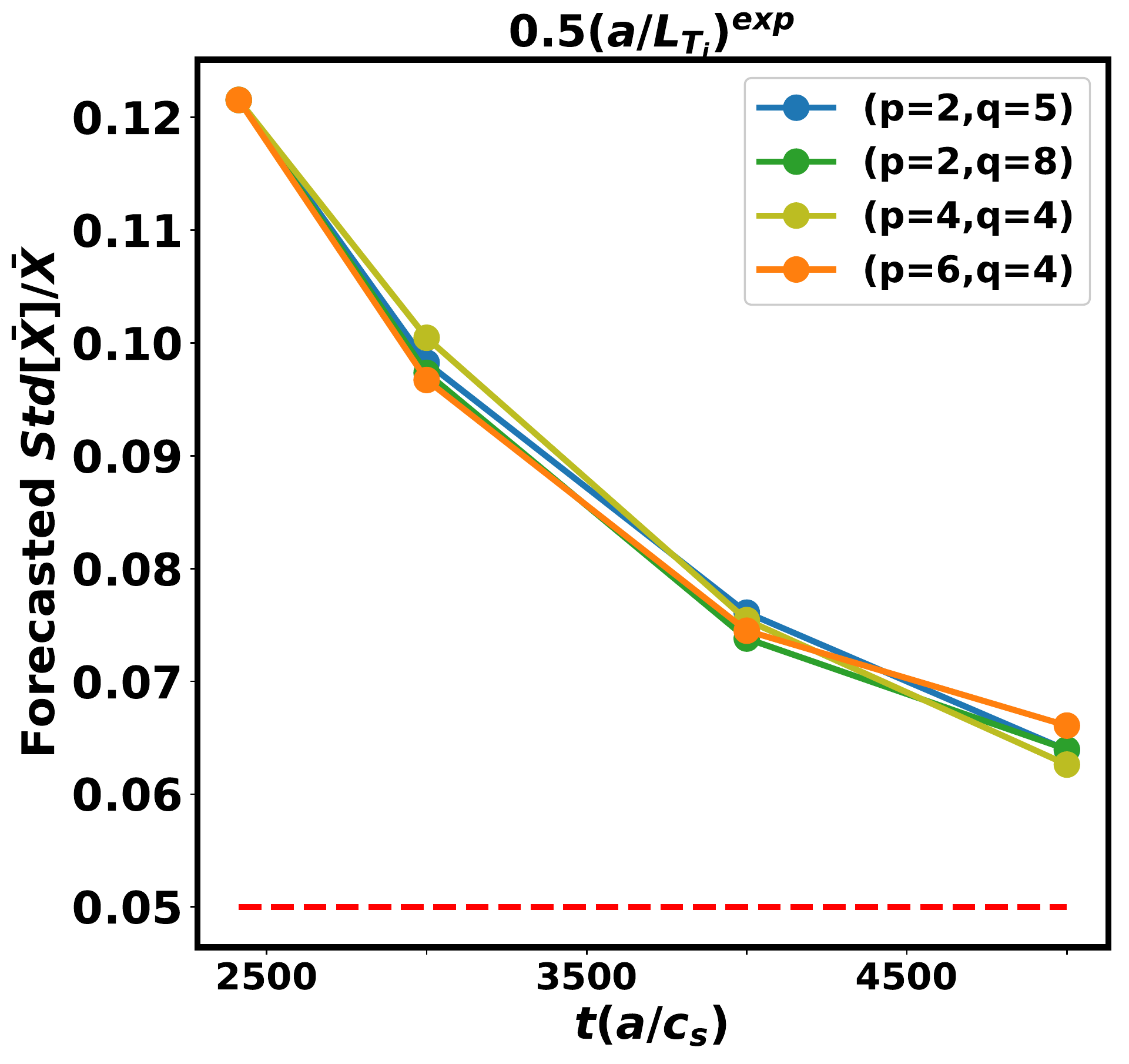}
  \end{minipage}
  \caption{Forecasting of ion energy flux fractional uncertainty at later simulation times for: (a) $a/L_{T_i} = (a/L_{T_i})^{exp}$, (b) $a/L_{T_i} = 0.8(a/L_{T_i})^{exp}$, (c) $a/L_{T_i} = 0.5(a/L_{T_i})^{exp}$. Red line shows a desired fractional uncertainty threshold of $5\%$.}
  \label{fig:armaforecast}
\end{figure*}

Once $p$, $q$, and $\sigma_\varepsilon$ is found, we can continue generating the time series up to a desired length, we use the inverse link $g^{-1}(\cdot)$ to transform back the the process to their original series, and perform sub-interval averaging of approximate link inverted ARMA fit to approximate the mean variance of simulation quantities at a later simulation time. In Fig. \ref{fig:armaforecast} we have forecasted the fractional uncertainty of ion energy flux for up $5000 (a/c_s)$ in each simulation case.

We observe as the simulation length increases the variance of the mean shrinks down. However, the variance reduction is not similar to treating the samples as uncorrelated, due to finite autocorrelation of lags. We can observe for $(a/L_{T_i})^{exp}$ simulation case, if one is aiming for five percent factional uncertainty, we tentatively need to run the simulation for up to $2000 (a/c_s)$. For the simulation case of $0.8(a/L_{T_i})^{exp}$, with current simulation length, the fractional uncertainty is at the desired level of five percent. For the $0.5(a/L_{T_i})^{exp}$ simulation case, we have forecasted the fractional uncertainty for all the four possible ARMA fits. We observe in the near marginal case the forecasted fractional uncertainty at $5000 (a/c_s)$ is between $6\%-7\%$, and simulation needs to be run even longer that $5000 (a/c_s)$ to achieve desired level of accuracy of less than five percent fractional uncertainty. {We should note that as we observed from our analytical analysis shown in Sec.} \ref{sec:armatest}{, the variance of the mean distribution of each ARMA model is a function of $\phi_i$ and $\theta_i$, thus the variance vary from one ARMA model to another one. Nonetheless, even for the non-converged $0.5(a/L_{T_i})^{exp}$ case, we shown that with multiple ARMA fits we can still forecast a range of fractional uncertainties at later simulation times.} These procedures can be used in future validation studies to help determine simulation length and computational resource requirements.

\section{Summary and Future Directions} \label{summary}

In this paper, we reviewed some previous approaches used within MFE community on estimating the variance of the mean distribution of simulated quantities. We compared the analytical mean variance of $ARMA(1,0)$ process with two previously used mean variance techniques, namely the integral correlation time, and the sub-interval averaging approaches. We found that the integral correlation time is very sensitive to the choice of kernel estimator, while sub-interval averaging of correlated measurements can be a robust method as long as we have enough temporal samples compare to the autocorrelation timescale. Moreover, we have studies fitting of the ARMA models to gyrokinetic simulated ion energy flux quantities. Through ARMA model fitting, we determined if the simulation has been run long enough, and forecasted the fractional uncertainty of turbulence quantity through later simulation times.

We should note that only ARMA models with normal error terms have been explored in this publication. Generalized ARMA models\cite{benjamin2003,zheng2015} can be studied and explored with different types of turbulence simulations, to provide a tool to model non-Gaussian processes, and acquire non-symmetric confidence intervals on the mean within nonlinear simulations. As MFE community further incorporating temporal UQ into plasma turbulence studies, more advanced methods such as Bayesian calibration of temporal models\cite{ling2013uncertainty} can also be seen as future avenue for stacking different types of uncertainties.

\begin{acknowledgments}
The authors would thank W. M. Nevins and D. R. Mikkelsen for their initial work on this topic, including implementation of the approaches described in Sec. \ref{sec:meanvaroldtechniques} into analysis software. C.H. also thanks both, as well as J. Parker, for many useful discussions on this topic. {Authors would also thank O. Meneghini for useful discussions.} The work was supported by the U.S. Department of Energy Office, Office of Sciences, Office of Fusion Energy Sciences award numbers DE-SC0006957 (Center for Simulation of Plasma Microturbulence) and DE-SC0018287 (AToM: Advanced Tokamak Modeling Environment). The simulations were performed on computing resources of the National Energy Research Scientific Computing Center (NERSC), a U.S. Department of Energy Office of Science User Facility operated under Contract No. DE-AC02-05CH11231.
\end{acknowledgments}

\bibliographystyle{apsrev4-1}
\bibliography{main}

\begin{thebibliography}{33}%
\makeatletter
\providecommand \@ifxundefined [1]{%
 \@ifx{#1\undefined}
}%
\providecommand \@ifnum [1]{%
 \ifnum #1\expandafter \@firstoftwo
 \else \expandafter \@secondoftwo
 \fi
}%
\providecommand \@ifx [1]{%
 \ifx #1\expandafter \@firstoftwo
 \else \expandafter \@secondoftwo
 \fi
}%
\providecommand \natexlab [1]{#1}%
\providecommand \enquote  [1]{``#1''}%
\providecommand \bibnamefont  [1]{#1}%
\providecommand \bibfnamefont [1]{#1}%
\providecommand \citenamefont [1]{#1}%
\providecommand \href@noop [0]{\@secondoftwo}%
\providecommand \href [0]{\begingroup \@sanitize@url \@href}%
\providecommand \@href[1]{\@@startlink{#1}\@@href}%
\providecommand \@@href[1]{\endgroup#1\@@endlink}%
\providecommand \@sanitize@url [0]{\catcode `\\12\catcode `\$12\catcode
  `\&12\catcode `\#12\catcode `\^12\catcode `\_12\catcode `\%12\relax}%
\providecommand \@@startlink[1]{}%
\providecommand \@@endlink[0]{}%
\providecommand \url  [0]{\begingroup\@sanitize@url \@url }%
\providecommand \@url [1]{\endgroup\@href {#1}{\urlprefix }}%
\providecommand \urlprefix  [0]{URL }%
\providecommand \Eprint [0]{\href }%
\providecommand \doibase [0]{http://dx.doi.org/}%
\providecommand \selectlanguage [0]{\@gobble}%
\providecommand \bibinfo  [0]{\@secondoftwo}%
\providecommand \bibfield  [0]{\@secondoftwo}%
\providecommand \translation [1]{[#1]}%
\providecommand \BibitemOpen [0]{}%
\providecommand \bibitemStop [0]{}%
\providecommand \bibitemNoStop [0]{.\EOS\space}%
\providecommand \EOS [0]{\spacefactor3000\relax}%
\providecommand \BibitemShut  [1]{\csname bibitem#1\endcsname}%
\let\auto@bib@innerbib\@empty
\bibitem [{\citenamefont {Holland}(2016)}]{holland2016}%
  \BibitemOpen
  \bibfield  {author} {\bibinfo {author} {\bibfnamefont {C.}~\bibnamefont
  {Holland}},\ }\href {\doibase 10.1063/1.4954151} {\bibfield  {journal}
  {\bibinfo  {journal} {Physics of Plasmas}\ }\textbf {\bibinfo {volume}
  {23}},\ \bibinfo {pages} {060901} (\bibinfo {year} {2016})}\BibitemShut
  {NoStop}%
\bibitem [{\citenamefont {Chilenski}\ \emph {et~al.}(2015)\citenamefont
  {Chilenski}, \citenamefont {Greenwald}, \citenamefont {Marzouk},
  \citenamefont {Howard}, \citenamefont {White}, \citenamefont {Rice},\ and\
  \citenamefont {Walk}}]{chilenski2015}%
  \BibitemOpen
  \bibfield  {author} {\bibinfo {author} {\bibfnamefont {M.}~\bibnamefont
  {Chilenski}}, \bibinfo {author} {\bibfnamefont {M.}~\bibnamefont
  {Greenwald}}, \bibinfo {author} {\bibfnamefont {Y.}~\bibnamefont {Marzouk}},
  \bibinfo {author} {\bibfnamefont {N.}~\bibnamefont {Howard}}, \bibinfo
  {author} {\bibfnamefont {A.}~\bibnamefont {White}}, \bibinfo {author}
  {\bibfnamefont {J.}~\bibnamefont {Rice}}, \ and\ \bibinfo {author}
  {\bibfnamefont {J.}~\bibnamefont {Walk}},\ }\href
  {http://stacks.iop.org/0029-5515/55/i=2/a=023012} {\bibfield  {journal}
  {\bibinfo  {journal} {Nuclear Fusion}\ }\textbf {\bibinfo {volume} {55}},\
  \bibinfo {pages} {023012} (\bibinfo {year} {2015})}\BibitemShut {NoStop}%
\bibitem [{\citenamefont {Vaezi}\ and\ \citenamefont
  {Holland}(2018)}]{vaezi2018}%
  \BibitemOpen
  \bibfield  {author} {\bibinfo {author} {\bibfnamefont {P.}~\bibnamefont
  {Vaezi}}\ and\ \bibinfo {author} {\bibfnamefont {C.}~\bibnamefont
  {Holland}},\ }\href {\doibase 10.1080/15361055.2017.1372987} {\bibfield
  {journal} {\bibinfo  {journal} {Fusion Science and Technology}\ }\textbf
  {\bibinfo {volume} {74}},\ \bibinfo {pages} {77} (\bibinfo {year}
  {2018})}\BibitemShut {NoStop}%
\bibitem [{\citenamefont {Vaezi}\ \emph {et~al.}(2018)\citenamefont {Vaezi},
  \citenamefont {Holland}, \citenamefont {Grierson}, \citenamefont {Staebler},
  \citenamefont {Smith},\ and\ \citenamefont {Meneghini}}]{vaezi2018b}%
  \BibitemOpen
  \bibfield  {author} {\bibinfo {author} {\bibfnamefont {P.}~\bibnamefont
  {Vaezi}}, \bibinfo {author} {\bibfnamefont {C.}~\bibnamefont {Holland}},
  \bibinfo {author} {\bibfnamefont {B.~A.}\ \bibnamefont {Grierson}}, \bibinfo
  {author} {\bibfnamefont {G.~M.}\ \bibnamefont {Staebler}}, \bibinfo {author}
  {\bibfnamefont {S.~P.}\ \bibnamefont {Smith}}, \ and\ \bibinfo {author}
  {\bibfnamefont {O.}~\bibnamefont {Meneghini}},\ }\href {\doibase
  10.1063/1.5053906} {\bibfield  {journal} {\bibinfo  {journal} {Physics of
  Plasmas}\ }\textbf {\bibinfo {volume} {25}},\ \bibinfo {pages} {102309}
  (\bibinfo {year} {2018})}\BibitemShut {NoStop}%
\bibitem [{\citenamefont {Staebler}\ \emph {et~al.}(2007)\citenamefont
  {Staebler}, \citenamefont {Kinsey},\ and\ \citenamefont
  {Waltz}}]{staebler2007}%
  \BibitemOpen
  \bibfield  {author} {\bibinfo {author} {\bibfnamefont {G.~M.}\ \bibnamefont
  {Staebler}}, \bibinfo {author} {\bibfnamefont {J.~E.}\ \bibnamefont
  {Kinsey}}, \ and\ \bibinfo {author} {\bibfnamefont {R.~E.}\ \bibnamefont
  {Waltz}},\ }\href {\doibase 10.1063/1.2436852} {\bibfield  {journal}
  {\bibinfo  {journal} {Physics of Plasmas}\ }\textbf {\bibinfo {volume}
  {14}},\ \bibinfo {pages} {055909} (\bibinfo {year} {2007})}\BibitemShut
  {NoStop}%
\bibitem [{\citenamefont {Mikkelsen}\ and\ \citenamefont
  {Dorland}(2005)}]{mikkelsen2008}%
  \BibitemOpen
  \bibfield  {author} {\bibinfo {author} {\bibfnamefont {D.~R.}\ \bibnamefont
  {Mikkelsen}}\ and\ \bibinfo {author} {\bibfnamefont {W.}~\bibnamefont
  {Dorland}},\ }\href@noop {} {\bibfield  {journal} {\bibinfo  {journal} {Bull.
  Am. Phys. Soc.}\ }\textbf {\bibinfo {volume} {50}},\ \bibinfo {pages} {196}
  (\bibinfo {year} {2005})}\BibitemShut {NoStop}%
\bibitem [{\citenamefont {Anderson}\ and\ \citenamefont
  {Hnat}(2017)}]{anderson2017}%
  \BibitemOpen
  \bibfield  {author} {\bibinfo {author} {\bibfnamefont {J.}~\bibnamefont
  {Anderson}}\ and\ \bibinfo {author} {\bibfnamefont {B.}~\bibnamefont
  {Hnat}},\ }\href {\doibase 10.1063/1.4984985} {\bibfield  {journal} {\bibinfo
   {journal} {Physics of Plasmas}\ }\textbf {\bibinfo {volume} {24}},\ \bibinfo
  {pages} {062301} (\bibinfo {year} {2017})}\BibitemShut {NoStop}%
\bibitem [{\citenamefont {Parker}\ \emph {et~al.}(2018)\citenamefont {Parker},
  \citenamefont {LoDestro},\ and\ \citenamefont {Campos}}]{parker2018}%
  \BibitemOpen
  \bibfield  {author} {\bibinfo {author} {\bibfnamefont {J.~B.}\ \bibnamefont
  {Parker}}, \bibinfo {author} {\bibfnamefont {L.~L.}\ \bibnamefont
  {LoDestro}}, \ and\ \bibinfo {author} {\bibfnamefont {A.}~\bibnamefont
  {Campos}},\ }\href {\doibase 10.3390/plasma1010012} {\bibfield  {journal}
  {\bibinfo  {journal} {Plasma}\ }\textbf {\bibinfo {volume} {1}},\ \bibinfo
  {pages} {126} (\bibinfo {year} {2018})}\BibitemShut {NoStop}%
\bibitem [{\citenamefont {Brown}\ \emph {et~al.}(1971)\citenamefont {Brown}
  \emph {et~al.}}]{brown1971}%
  \BibitemOpen
  \bibfield  {author} {\bibinfo {author} {\bibfnamefont {B.~M.}\ \bibnamefont
  {Brown}} \emph {et~al.},\ }\href@noop {} {\bibfield  {journal} {\bibinfo
  {journal} {The Annals of Mathematical Statistics}\ }\textbf {\bibinfo
  {volume} {42}},\ \bibinfo {pages} {59} (\bibinfo {year} {1971})}\BibitemShut
  {NoStop}%
\bibitem [{\citenamefont {Zhang}(2006)}]{zhang2006}%
  \BibitemOpen
  \bibfield  {author} {\bibinfo {author} {\bibfnamefont {N.~F.}\ \bibnamefont
  {Zhang}},\ }\href {http://stacks.iop.org/0026-1394/43/i=4/a=S15} {\bibfield
  {journal} {\bibinfo  {journal} {Metrologia}\ }\textbf {\bibinfo {volume}
  {43}},\ \bibinfo {pages} {S276} (\bibinfo {year} {2006})}\BibitemShut
  {NoStop}%
\bibitem [{\citenamefont {Andrews}(1991)}]{andrews1991}%
  \BibitemOpen
  \bibfield  {author} {\bibinfo {author} {\bibfnamefont {D.~W.~K.}\
  \bibnamefont {Andrews}},\ }\href {http://www.jstor.org/stable/2938229}
  {\bibfield  {journal} {\bibinfo  {journal} {Econometrica}\ }\textbf {\bibinfo
  {volume} {59}},\ \bibinfo {pages} {817} (\bibinfo {year} {1991})}\BibitemShut
  {NoStop}%
\bibitem [{\citenamefont {Anderson}(2011)}]{anderson2011}%
  \BibitemOpen
  \bibfield  {author} {\bibinfo {author} {\bibfnamefont {T.~W.}\ \bibnamefont
  {Anderson}},\ }\href@noop {} {\emph {\bibinfo {title} {The statistical
  analysis of time series}}},\ Vol.~\bibinfo {volume} {19}\ (\bibinfo
  {publisher} {John Wiley \& Sons},\ \bibinfo {year} {2011})\BibitemShut
  {NoStop}%
\bibitem [{\citenamefont {Nevins}(2004)}]{nevins2004}%
  \BibitemOpen
  \bibfield  {author} {\bibinfo {author} {\bibfnamefont {W.~M.}\ \bibnamefont
  {Nevins}},\ }\href@noop {} {\emph {\bibinfo {title} {GKV User Manual}}},\
  \bibinfo {type} {Tech. Rep.}\ \bibinfo {number} {UCRL-TR-206016}\ (\bibinfo
  {institution} {Lawrence Livermore National Laboratory},\ \bibinfo {year}
  {2004})\BibitemShut {NoStop}%
\bibitem [{\citenamefont {Mikkelsen}\ and\ \citenamefont
  {Dorland}(2008)}]{mikkelsen2008b}%
  \BibitemOpen
  \bibfield  {author} {\bibinfo {author} {\bibfnamefont {D.~R.}\ \bibnamefont
  {Mikkelsen}}\ and\ \bibinfo {author} {\bibfnamefont {W.}~\bibnamefont
  {Dorland}},\ }\href {\doibase 10.1103/PhysRevLett.101.135003} {\bibfield
  {journal} {\bibinfo  {journal} {Phys. Rev. Lett.}\ }\textbf {\bibinfo
  {volume} {101}},\ \bibinfo {pages} {135003} (\bibinfo {year}
  {2008})}\BibitemShut {NoStop}%
\bibitem [{\citenamefont {Parzen}(1963)}]{parzen1963}%
  \BibitemOpen
  \bibfield  {author} {\bibinfo {author} {\bibfnamefont {E.}~\bibnamefont
  {Parzen}},\ }\href {http://www.jstor.org/stable/25049287} {\bibfield
  {journal} {\bibinfo  {journal} {Sankhy: The Indian Journal of Statistics,
  Series A (1961-2002)}\ }\textbf {\bibinfo {volume} {25}},\ \bibinfo {pages}
  {383} (\bibinfo {year} {1963})}\BibitemShut {NoStop}%
\bibitem [{\citenamefont {Choi}(2012)}]{choi2012arma}%
  \BibitemOpen
  \bibfield  {author} {\bibinfo {author} {\bibfnamefont {B.}~\bibnamefont
  {Choi}},\ }\href@noop {} {\emph {\bibinfo {title} {ARMA model
  identification}}}\ (\bibinfo  {publisher} {Springer Science \& Business
  Media},\ \bibinfo {year} {2012})\BibitemShut {NoStop}%
\bibitem [{\citenamefont {Crack}\ and\ \citenamefont
  {Ledoit}(2004)}]{crack2004}%
  \BibitemOpen
  \bibfield  {author} {\bibinfo {author} {\bibfnamefont {T.~F.}\ \bibnamefont
  {Crack}}\ and\ \bibinfo {author} {\bibfnamefont {O.}~\bibnamefont {Ledoit}},\
  }\href@noop {} {\bibfield  {journal} {\bibinfo  {journal} {Social Science
  Research Network}\ ,\ \bibinfo {pages} {1}} (\bibinfo {year}
  {2004})}\BibitemShut {NoStop}%
\bibitem [{\citenamefont {Thompson}(2010)}]{thompson2010comparison}%
  \BibitemOpen
  \bibfield  {author} {\bibinfo {author} {\bibfnamefont {M.~B.}\ \bibnamefont
  {Thompson}},\ }\href {https://arxiv.org/abs/1011.0175} {\bibfield  {journal}
  {\bibinfo  {journal} {arXiv preprint arXiv:1011.0175}\ } (\bibinfo {year}
  {2010})}\BibitemShut {NoStop}%
\bibitem [{\citenamefont {Luce}\ \emph {et~al.}(2018)\citenamefont {Luce},
  \citenamefont {Burrell}, \citenamefont {Holland}, \citenamefont {Marinoni},
  \citenamefont {Petty}, \citenamefont {Smith}, \citenamefont {Austin},
  \citenamefont {Grierson},\ and\ \citenamefont {Zeng}}]{luce2017}%
  \BibitemOpen
  \bibfield  {author} {\bibinfo {author} {\bibfnamefont {T.}~\bibnamefont
  {Luce}}, \bibinfo {author} {\bibfnamefont {K.}~\bibnamefont {Burrell}},
  \bibinfo {author} {\bibfnamefont {C.}~\bibnamefont {Holland}}, \bibinfo
  {author} {\bibfnamefont {A.}~\bibnamefont {Marinoni}}, \bibinfo {author}
  {\bibfnamefont {C.}~\bibnamefont {Petty}}, \bibinfo {author} {\bibfnamefont
  {S.}~\bibnamefont {Smith}}, \bibinfo {author} {\bibfnamefont
  {M.}~\bibnamefont {Austin}}, \bibinfo {author} {\bibfnamefont
  {B.}~\bibnamefont {Grierson}}, \ and\ \bibinfo {author} {\bibfnamefont
  {L.}~\bibnamefont {Zeng}},\ }\href
  {http://stacks.iop.org/0029-5515/58/i=2/a=026023} {\bibfield  {journal}
  {\bibinfo  {journal} {Nuclear Fusion}\ }\textbf {\bibinfo {volume} {58}},\
  \bibinfo {pages} {026023} (\bibinfo {year} {2018})}\BibitemShut {NoStop}%
\bibitem [{\citenamefont {Candy}\ \emph {et~al.}(2016)\citenamefont {Candy},
  \citenamefont {Belli},\ and\ \citenamefont {Bravenec}}]{candy2016}%
  \BibitemOpen
  \bibfield  {author} {\bibinfo {author} {\bibfnamefont {J.}~\bibnamefont
  {Candy}}, \bibinfo {author} {\bibfnamefont {E.}~\bibnamefont {Belli}}, \ and\
  \bibinfo {author} {\bibfnamefont {R.}~\bibnamefont {Bravenec}},\ }\href
  {\doibase https://doi.org/10.1016/j.jcp.2016.07.039} {\bibfield  {journal}
  {\bibinfo  {journal} {Journal of Computational Physics}\ }\textbf {\bibinfo
  {volume} {324}},\ \bibinfo {pages} {73 } (\bibinfo {year}
  {2016})}\BibitemShut {NoStop}%
\bibitem [{\citenamefont {Belli}\ and\ \citenamefont
  {Candy}(2008)}]{belli2008}%
  \BibitemOpen
  \bibfield  {author} {\bibinfo {author} {\bibfnamefont {E.~A.}\ \bibnamefont
  {Belli}}\ and\ \bibinfo {author} {\bibfnamefont {J.}~\bibnamefont {Candy}},\
  }\href {http://stacks.iop.org/0741-3335/50/i=9/a=095010} {\bibfield
  {journal} {\bibinfo  {journal} {Plasma Physics and Controlled Fusion}\
  }\textbf {\bibinfo {volume} {50}},\ \bibinfo {pages} {095010} (\bibinfo
  {year} {2008})}\BibitemShut {NoStop}%
\bibitem [{\citenamefont {Belli}\ and\ \citenamefont
  {Candy}(2012)}]{belli2012}%
  \BibitemOpen
  \bibfield  {author} {\bibinfo {author} {\bibfnamefont {E.~A.}\ \bibnamefont
  {Belli}}\ and\ \bibinfo {author} {\bibfnamefont {J.}~\bibnamefont {Candy}},\
  }\href {http://stacks.iop.org/0741-3335/54/i=1/a=015015} {\bibfield
  {journal} {\bibinfo  {journal} {Plasma Physics and Controlled Fusion}\
  }\textbf {\bibinfo {volume} {54}},\ \bibinfo {pages} {015015} (\bibinfo
  {year} {2012})}\BibitemShut {NoStop}%
\bibitem [{\citenamefont {Miller}\ \emph {et~al.}(1998)\citenamefont {Miller},
  \citenamefont {Chu}, \citenamefont {Greene}, \citenamefont {Lin-Liu},\ and\
  \citenamefont {Waltz}}]{miller1998}%
  \BibitemOpen
  \bibfield  {author} {\bibinfo {author} {\bibfnamefont {R.~L.}\ \bibnamefont
  {Miller}}, \bibinfo {author} {\bibfnamefont {M.~S.}\ \bibnamefont {Chu}},
  \bibinfo {author} {\bibfnamefont {J.~M.}\ \bibnamefont {Greene}}, \bibinfo
  {author} {\bibfnamefont {Y.~R.}\ \bibnamefont {Lin-Liu}}, \ and\ \bibinfo
  {author} {\bibfnamefont {R.~E.}\ \bibnamefont {Waltz}},\ }\href {\doibase
  10.1063/1.872666} {\bibfield  {journal} {\bibinfo  {journal} {Physics of
  Plasmas}\ }\textbf {\bibinfo {volume} {5}},\ \bibinfo {pages} {973} (\bibinfo
  {year} {1998})}\BibitemShut {NoStop}%
\bibitem [{\citenamefont {Candy}\ \emph {et~al.}(2009)\citenamefont {Candy},
  \citenamefont {Holland}, \citenamefont {Waltz}, \citenamefont {Fahey},\ and\
  \citenamefont {Belli}}]{candy2009}%
  \BibitemOpen
  \bibfield  {author} {\bibinfo {author} {\bibfnamefont {J.}~\bibnamefont
  {Candy}}, \bibinfo {author} {\bibfnamefont {C.}~\bibnamefont {Holland}},
  \bibinfo {author} {\bibfnamefont {R.~E.}\ \bibnamefont {Waltz}}, \bibinfo
  {author} {\bibfnamefont {M.~R.}\ \bibnamefont {Fahey}}, \ and\ \bibinfo
  {author} {\bibfnamefont {E.}~\bibnamefont {Belli}},\ }\href {\doibase
  10.1063/1.3167820} {\bibfield  {journal} {\bibinfo  {journal} {Physics of
  Plasmas}\ }\textbf {\bibinfo {volume} {16}},\ \bibinfo {pages} {060704}
  (\bibinfo {year} {2009})}\BibitemShut {NoStop}%
\bibitem [{\citenamefont {Candy}\ and\ \citenamefont
  {Belli}(2018)}]{candy2018}%
  \BibitemOpen
  \bibfield  {author} {\bibinfo {author} {\bibfnamefont {J.}~\bibnamefont
  {Candy}}\ and\ \bibinfo {author} {\bibfnamefont {E.}~\bibnamefont {Belli}},\
  }\href {\doibase https://doi.org/10.1016/j.jcp.2017.12.020} {\bibfield
  {journal} {\bibinfo  {journal} {Journal of Computational Physics}\ }\textbf
  {\bibinfo {volume} {356}},\ \bibinfo {pages} {448 } (\bibinfo {year}
  {2018})}\BibitemShut {NoStop}%
\bibitem [{\citenamefont {Box}\ \emph {et~al.}(2015)\citenamefont {Box},
  \citenamefont {Jenkins}, \citenamefont {Reinsel},\ and\ \citenamefont
  {Ljung}}]{box2015time}%
  \BibitemOpen
  \bibfield  {author} {\bibinfo {author} {\bibfnamefont {G.~E.}\ \bibnamefont
  {Box}}, \bibinfo {author} {\bibfnamefont {G.~M.}\ \bibnamefont {Jenkins}},
  \bibinfo {author} {\bibfnamefont {G.~C.}\ \bibnamefont {Reinsel}}, \ and\
  \bibinfo {author} {\bibfnamefont {G.~M.}\ \bibnamefont {Ljung}},\ }\href@noop
  {} {\emph {\bibinfo {title} {Time series analysis: forecasting and
  control}}}\ (\bibinfo  {publisher} {John Wiley \& Sons},\ \bibinfo {year}
  {2015})\BibitemShut {NoStop}%
\bibitem [{\citenamefont {Sakia}(1992)}]{boxcox1992}%
  \BibitemOpen
  \bibfield  {author} {\bibinfo {author} {\bibfnamefont {R.~M.}\ \bibnamefont
  {Sakia}},\ }\href {http://www.jstor.org/stable/2348250} {\bibfield  {journal}
  {\bibinfo  {journal} {Journal of the Royal Statistical Society. Series D (The
  Statistician)}\ }\textbf {\bibinfo {volume} {41}},\ \bibinfo {pages} {169}
  (\bibinfo {year} {1992})}\BibitemShut {NoStop}%
\bibitem [{\citenamefont {Benjamin}\ \emph {et~al.}(2003)\citenamefont
  {Benjamin}, \citenamefont {Rigby},\ and\ \citenamefont
  {Stasinopoulos}}]{benjamin2003}%
  \BibitemOpen
  \bibfield  {author} {\bibinfo {author} {\bibfnamefont {M.~A.}\ \bibnamefont
  {Benjamin}}, \bibinfo {author} {\bibfnamefont {R.~A.}\ \bibnamefont {Rigby}},
  \ and\ \bibinfo {author} {\bibfnamefont {D.~M.}\ \bibnamefont
  {Stasinopoulos}},\ }\href {\doibase 10.1198/016214503388619238} {\bibfield
  {journal} {\bibinfo  {journal} {Journal of the American Statistical
  Association}\ }\textbf {\bibinfo {volume} {98}},\ \bibinfo {pages} {214}
  (\bibinfo {year} {2003})}\BibitemShut {NoStop}%
\bibitem [{\citenamefont {Zheng}\ \emph {et~al.}(2015)\citenamefont {Zheng},
  \citenamefont {Xiao},\ and\ \citenamefont {Chen}}]{zheng2015}%
  \BibitemOpen
  \bibfield  {author} {\bibinfo {author} {\bibfnamefont {T.}~\bibnamefont
  {Zheng}}, \bibinfo {author} {\bibfnamefont {H.}~\bibnamefont {Xiao}}, \ and\
  \bibinfo {author} {\bibfnamefont {R.}~\bibnamefont {Chen}},\ }\href {\doibase
  https://doi.org/10.1016/j.jeconom.2015.03.040} {\bibfield  {journal}
  {\bibinfo  {journal} {Journal of Econometrics}\ }\textbf {\bibinfo {volume}
  {189}},\ \bibinfo {pages} {492 } (\bibinfo {year} {2015})},\ \bibinfo {note}
  {frontiers in Time Series and Financial Econometrics}\BibitemShut {NoStop}%
\bibitem [{\citenamefont {Hannan}\ and\ \citenamefont
  {Rissanen}(1982)}]{hannan1982}%
  \BibitemOpen
  \bibfield  {author} {\bibinfo {author} {\bibfnamefont {E.~J.}\ \bibnamefont
  {Hannan}}\ and\ \bibinfo {author} {\bibfnamefont {J.}~\bibnamefont
  {Rissanen}},\ }\href {\doibase 10.1093/biomet/69.1.81} {\bibfield  {journal}
  {\bibinfo  {journal} {Biometrika}\ }\textbf {\bibinfo {volume} {69}},\
  \bibinfo {pages} {81} (\bibinfo {year} {1982})}\BibitemShut {NoStop}%
\bibitem [{\citenamefont {D'Agostino}\ and\ \citenamefont
  {Pearson}(1973)}]{dagostino1973}%
  \BibitemOpen
  \bibfield  {author} {\bibinfo {author} {\bibfnamefont {R.}~\bibnamefont
  {D'Agostino}}\ and\ \bibinfo {author} {\bibfnamefont {E.~S.}\ \bibnamefont
  {Pearson}},\ }\href {\doibase 10.1093/biomet/60.3.613} {\bibfield  {journal}
  {\bibinfo  {journal} {Biometrika}\ }\textbf {\bibinfo {volume} {60}},\
  \bibinfo {pages} {613} (\bibinfo {year} {1973})}\BibitemShut {NoStop}%
\bibitem [{\citenamefont {Ljung}\ and\ \citenamefont {Box}(1978)}]{ljung1978}%
  \BibitemOpen
  \bibfield  {author} {\bibinfo {author} {\bibfnamefont {G.~M.}\ \bibnamefont
  {Ljung}}\ and\ \bibinfo {author} {\bibfnamefont {G.~E.~P.}\ \bibnamefont
  {Box}},\ }\href {\doibase 10.1093/biomet/65.2.297} {\bibfield  {journal}
  {\bibinfo  {journal} {Biometrika}\ }\textbf {\bibinfo {volume} {65}},\
  \bibinfo {pages} {297} (\bibinfo {year} {1978})}\BibitemShut {NoStop}%
\bibitem [{\citenamefont {Ling}(2013)}]{ling2013uncertainty}%
  \BibitemOpen
  \bibfield  {author} {\bibinfo {author} {\bibfnamefont {Y.}~\bibnamefont
  {Ling}},\ }\href@noop {} {\emph {\bibinfo {title} {Uncertainty quantification
  in time-dependent reliability analysis}}}\ (\bibinfo  {publisher} {Vanderbilt
  University},\ \bibinfo {year} {2013})\BibitemShut {NoStop}%
\end{thebibliography}%

\end{document}